\documentclass[pra,aps,twocolumn,superscriptaddress,showpacs]{revtex4}

\usepackage{amsmath}
\usepackage{amssymb}
\usepackage{graphicx}
\usepackage{float}
\usepackage{bbm}
\usepackage{color}

\newcommand{\dx}[0]{{\rm d}\!x}
\newcommand{\ddx}[0]{{\rm d}^3\!x}
\newcommand{\dt}[0]{{\rm d}t}

\newcommand{\rmi}{{\rm i}}
\newcommand{\rme}{{\rm e}}
\newcommand{\rmd}{{\rm d}}
\newcommand{\tr}[1]{{\rm Tr_R}\left\{#1\right\}}
\newcommand{\EV}[1]{\left\langle #1 \right\rangle}

\newcommand{\ket}[1]{\left | \, #1 \right \rangle}
\newcommand{\bra}[1]{\left \langle #1 \, \right |}
\newcommand{\bx}[0]{{\bf x}}
\newcommand{\bn}[0]{{\bf n}}
\newcommand{\bm}[0]{{\bf m}}
\newcommand{\bk}[0]{{\bf k}}

\newcommand{\dq}[0]{{\delta \!q}}

\newcommand{\ps}[1]{\hat\psi_{#1}(\bx)}
\newcommand{\psd}[1]{\hat\psi^\dagger_{#1}(\bx)}

\newcommand{\rmm}[0]{{\rm m}}
\newcommand{\be}[0]{{\bf e}}
\newcommand{\A}[2]{\hat A_{#1}^{#2}}
\newcommand{\Ad}[2]{\left(\hat A_{#1}^{#2}\right)^\dagger}
\newcommand{\ketm}[0]{\left| \bm \right\rangle}
\newcommand{\bram}[0]{\left\langle \bm\right|}
\newcommand{\ketn}[0]{\left| \bn \right\rangle}
\newcommand{\bran}[0]{\left\langle \bn \right|}

\newcommand{\half}{\mbox{$\textstyle \frac{1}{2}$}}
\newcommand{\braket}[2]{\left\langle\, #1\,|\,#2\,\right\rangle}

\newcommand{\av}[1]{\langle #1\rangle}

\begin{document}
\title{
Dissipative dynamics of atomic Hubbard models coupled to a phonon bath:\newline
Dark state cooling of atoms within a Bloch band of an optical lattice}

\author{A. Griessner}
\affiliation{Institute for Quantum Optics and Quantum Information of the Austrian Academy of
Sciences, A-6020 Innsbruck, Austria} \affiliation{Institute for Theoretical Physics, University of
Innsbruck, A-6020 Innsbruck, Austria}
\author{A. J. Daley}
\affiliation{Institute for Quantum Optics and Quantum Information of the Austrian Academy of
Sciences, A-6020 Innsbruck, Austria} \affiliation{Institute for Theoretical Physics, University of
Innsbruck, A-6020 Innsbruck, Austria}
\author{S. R. Clark}
\affiliation{Clarendon Laboratory, University of Oxford, Parks Road, Oxford OX1 3PU, United
Kingdom}
\author{D. Jaksch}
\affiliation{Clarendon Laboratory, University of Oxford, Parks Road, Oxford OX1 3PU, United
Kingdom}
\author{P. Zoller}
\affiliation{Institute for Quantum Optics and Quantum Information of the Austrian Academy of
Sciences, A-6020 Innsbruck, Austria} \affiliation{Institute for Theoretical Physics, University of
Innsbruck, A-6020 Innsbruck, Austria}

\begin{abstract}We analyse a laser assisted sympathetic cooling scheme for atoms within the lowest
Bloch band of an optical lattice. This scheme borrows ideas from sub-recoil laser cooling,
implementing them in a new context in which the atoms in the lattice are coupled to a BEC
reservoir. In this scheme, excitation of atoms between Bloch bands replaces the internal structure
of atoms in normal laser cooling, and spontaneous emission of photons is replaced by creation of
excitations in the BEC reservoir. We analyse the cooling process for many bosons and fermions, and
obtain possible temperatures corresponding to a small fraction of the Bloch band width within our
model. This system can be seen as a novel realisation of a many-body open quantum system.
\end{abstract}

\date{December 11, 2006}
\pacs{03.75.Lm, 32.80.Pj, 42.50.-p} \maketitle


\section{Introduction}

\label{Section:Introduction}
 New frontiers in atomic physics have often been enabled by
the development of new cooling techniques. Examples are provided by laser
cooling and evaporative cooling \cite{LCMetcalf, LCRMPChu, LCRMPCohen,
LCRMPPhillips}, which underly the exciting experimental advances to realize
Bose Einstein condensates (BEC) and quantum degenerate Fermi gases of atoms and
molecules \cite{NatureInsight,StringariPitaevski,PethickSmith, FermiJin,
FermiInguscio, FermiThomas, FermiHulet, MoleculeJin, MoleculeGrimm,
MoleculeKetterle, MoleculeSalomon}. Recently, quantum degenerate gases of
bosons and fermions have been loaded into optical lattices in one, two and
three-dimensional configurations \cite{OLDJ,OLPZFermions, OLBloch, OLEsslinger,
OLSengstock, OLKetterle, OLPhillips, OLInguscio, OLDenschlag, OLRempe}. This
makes it possible to realize atomic Hubbard dynamics with controllable
parameters, opening the door to the study (and simulation) of strongly
correlated systems with cold atoms. Control via external fields allows the
engineering of atomic lattice Hamiltonians for boson, fermion and spin models
\cite{reviewPZ,reviewLewenstein, reviewBloch}, which for a long time have been
the focus of research in theoretical condensed matter physics. However, one of
the most challenging obstacles in the realization of some of the most
interesting condensed matter systems in current experiments is the need for
lower temperatures of the atoms {\it within a Bloch band} of the optical
lattice
\footnote{ For example, the realization of a d-wave superconductor requires
temperatures on the order of $4\%$ of the Bloch band width (for details see S.
Trebst, U. Schollw\"{o}ck, M. Troyer, and P. Zoller, Phys. Rev. Lett.
\textbf{96}, 250402 (2006)).}.

In this paper we will analyze a configuration where atoms moving in the lowest Bloch band of an
optical lattice are cooled via {\em laser assisted} sympathetic cooling with a heatbath
represented by a BEC of atoms. A unique feature of the present scheme is that the achievable
temperatures of the atoms within a single Bloch band are significantly {\em lower} than those of
the cooling reservoir, reaching a temperature of a small fraction of the Bloch band width. From a
physics point of view, a guiding idea of the present work is formal analogies with {\em laser
cooling} \cite{LCMetcalf, LCRMPChu, LCRMPCohen, LCRMPPhillips}. In laser cooling the motional
degrees of an atom are coupled via laser excitation of the electrons to the effective zero
temperature (photon) reservoir, i.e., the vacuum modes of the optical light field. The spontaneous
emission of photons carries away the entropy allowing a purification (cooling) of the motional
state of the atom. In the present work we will follow a similar scenario by coupling atoms moving
in the lowest Bloch band via laser assisted processes to an excited band, which can decay back to
the lowest band by spontaneous emission of phonon (or particle-like) excitations in the BEC. We
note that this cooling process is atom number conserving, in contrast to filtering and evaporative
cooling technquies \cite{PethickSmith, filteringPäda,filteringCirac}. The present paper focuses on
a specific protocol for cooling within a Bloch band which follows analogies with (subrecoil) Raman
laser cooling, as developed in seminal work by Kasevich and Chu \cite{raman_chu}.

We see an important aspect of the present work in pointing out the formal
analogies between open quantum systems familiar from quantum optics with light
fields and the present setup of atoms in optical lattices coupled to a phonon
bath. These analogies not only stimulate the transfer of well established ideas
of, e.g., laser cooling, to a new context, but also provides a new realization
of an open quantum optical system with dissipative dynamics of cold atoms in
optical lattices. The present paper extends and gives details of the analysis
of work published in Ref. \cite{CoolingPRL}, in which setup we assume a
homogeneous optical lattice, without an additional trapping potential for
lattice atoms. As an additional remark, we also present an extension of these
ideas to a form of spatial sideband cooling in a harmonic trapping potential.

\begin{center}
    \begin{figure}[htp]
        \includegraphics{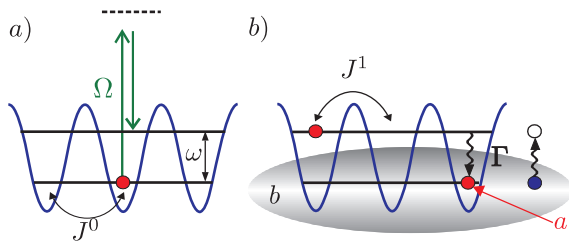}\caption{Setup for the Raman cooling scheme in
the optical lattice. a) A Raman laser setup couples atoms in the lowest Bloch band to the first
excited band. b) The lattice is immersed into a BEC of species $b$, which serves as a cold
reservoir for the lattice atoms, leading to decay of lattice atoms in an excited motional state
back to the lowest band via collisional interaction with the reservoir and the creation of
Bogoliubov excitations in the BEC (as sketched in the far right of the figure).}\label{Fig:Setup}
    \end{figure}
\end{center}

We consider a setup as shown in Fig.~\ref{Fig:Setup}, where the internal electronic states of
atoms familiar from laser cooling, are replaced with the two lowest Bloch bands of an optical
lattice, trapping atoms of a species $a$. In analogy to laser cooling the energy of particles in
the lowest Bloch band is upconverted by transferring lattice atoms to the first excited band via a
Raman transition. We immerse the atoms in the optical lattice in a BEC \cite{AJcooling}, which
serves as a reservoir for the lattice atoms, and should be seen as the counterpart of the vacuum
modes of the radiation field in laser cooling, carrying away motional energy from the system. In
our scheme, cooling is achieved in two steps: (i) We design a sequence of excitation pulses which
efficiently excites atoms with high quasi-momenta in the lattice, but do not couple to atoms with
$q=0$. (ii) In a second step, the coupling to the BEC reservoir leads to decay of excited lattice
atoms back to the ground state via emission of a Bogoliubov excitation into the BEC, thereby
randomizing the quasi-momentum. Repeating these two processes leads to an accumulation of the
atoms in a narrow region around $q=0$, i.e., to cooling, in analogy to the Kasevich-Chu scheme of
Raman laser cooling for free atoms \cite{VSCPT,raman_chu}. We will show below, that our method
works away from the limit of unit filling of the lattice, and is capable of cooling single atoms
and many non-interacting bosons and fermions. The method can be utilized in a scheme to create
strongly interacting phases of atoms in the optical lattice, by first cooling non-interacting
atoms (thereby exploiting the tunability of interactions, e.g., via Feshbach resonances) and
subsequently ramping up the interaction.

The paper is organized as follows. In the following section \ref{Section:LaserCooling} we will
present a short overview of laser cooling as is relevant for understanding the physical analogies
with our cooling method outlined in later sections. Sec.~\ref{Section:Idea} presents our model. We
will then illustrate the details of the cooling protocol for the case of a single particle in
Sec.~\ref{Section:Single_Particle}. In Sec.~\ref{Section:NParticles} we will discuss how the
system can be adapted to achieve cooling for many non-interacting bosons and fermions and we
investigate the affects of interaction on the cooling scheme for many bosons. In
Sec.~\ref{Section:ramping} we will show how the interaction strength between already cooled
lattice atoms can be ramped up to achieve cold strongly correlated gases. In
Sec.~\ref{Section:compacting} we discuss how similar ideas could be used to compact Fermions in a
harmonic trap, and we then conclude and summarize the ideas and main results in
Sec.~\ref{Section:Conclusion}.


\section{Laser Cooling}\label{Section:LaserCooling}

In this section we will outline the main ingredients needed for the description of laser cooling
and give a short overview of the development of the different laser cooling schemes. The structure
presented here is directly related to the cooling scheme we present in Sec.~\ref{Section:Idea}, as
are the concepts of subrecoil laser cooling, which we discuss in more detail at the end of this
section.

As described in the introduction, in a laser cooling scheme, an atom interacts with a laser and is
coupled to the electromagnetic radiation field, i.e., one considers a system of the form
\begin{align}
  \hat H=\hat H_a+ \hat H_b+ \hat H_{\rm int}+ \hat H_{\rm LS}. \label{HLC}
\end{align} Here, the atomic Hamiltonian $\hat H_a$, includes the atom's motion, possibly an
external trapping potential and the internal structure, usually modelled as a two or a three level
system. The atom interacts with the electromagnetic radiation field, described by $\hat H_b$, and
this interaction is described by $\hat H_{\rm int}$. In addition, a laser setup couples the atom's
internal states, which is described by $\hat H_{\rm LS}$.

The two steps of laser cooling, i.e., the upconversion of the energy due to the
laser and the removal of system energy due to spontaneous emission, can be
described in terms of a master equation in the Born-Markov approximation for
the reduced system density operator $\hat\rho$ (see, e.g.,
{\cite{LCstandingwave}),
\begin{align}
  &{\dot{\hat \rho}} =\rmi [\hat H_a+ \hat H_{LS},\hat\rho]-\frac{\Gamma}{2}
  ( \hat \sigma^+ \hat \sigma^- \hat \rho +\hat \rho
\hat \sigma^+ \hat\sigma^-)\nonumber\\
&\qquad+\Gamma\int_{-1}^{+1} \rmd u N(u) \rme^{\rmi k_l \hat x u} \hat \sigma^-
\hat \rho \hat \sigma^+ \rme^{-\rmi k_l \hat x u},\label{MELaserCooling}
\end{align} where we have specialized to atomic motion in one dimension and to
a two level system for sake of simplicity. The atom's position operator is
denoted $\hat x$, $\hat \sigma^+$ ($\hat \sigma^-$) is the raising (lowering)
operator corresponding to the electronic transition, $\Gamma$ denotes the
linewidth of the excited state, and $k_l$ the wavenumber of the laser light.
Note that photons can be spontaneously emitted in all three dimensions, and the
angular dependence of the spontaneous emission is accounted for by the
normalized dipole distribution $N(u)$ in this one dimensional model.

These basic ingredients have been utilized in the development of various
different laser cooling schemes over the last years. This began with Doppler
cooling \cite{LCMetcalf}, where two counter-propagating laser beams incident on
an atom, so that the radiation pressure of the two beams compensates for an
atom at rest, but is Doppler shifted towards resonance and thus enhanced for
the counter-propagating beam in the case of a moving atom. In Doppler cooling
the final temperature is limited by the linewidth of the excited state. Lower
temperatures were obtained in various schemes, including polarisation gradient
cooling and Sisyphus cooling \cite{coolingbelowTD}, where the limiting
temperature is given by the recoil energy an atom receives during the emission
of a single photon. Cooling of atoms even below this single photon recoil limit
was proposed and observed in the form of Raman laser cooling and velocity
selective coherent population trapping \cite{VSCPT,raman_chu}.

\subsection{Subrecoil Laser Cooling}

The basic idea in subrecoil laser cooling is to make the photon absorption rate velocity dependent
and, in particular, vanishing for a {\em dark state}. The atoms will consequently accumulate in
this state during the cooling process. In the following we will describe the examples of Raman
cooling and velocity selective coherent population trapping (VSCPT).

\subsubsection{Raman Cooling}

\begin{center}
  \begin{figure}[htp]
    \includegraphics{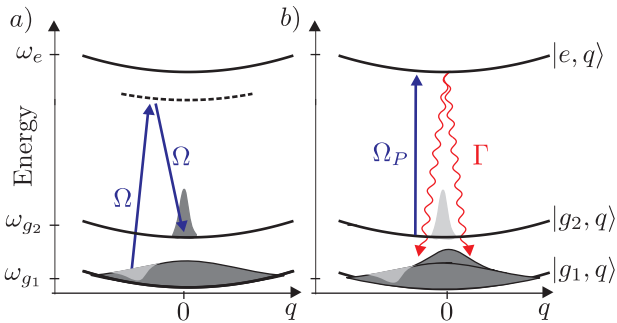}
    \caption{Schematic picture of one excitation and decay step in subrecoil
Raman laser cooling. a) Atoms from a region with high momentum $|q|>0$ are transferred from one
ground state $g_1$ to a second ground state $g_2$. b) The atoms in state $g_2$ are optically
pumped back to the initial state $g_1$ via the electronically excited state $e$, thereby
randomizing the momentum due to the recoil kick. Cooling is achieved by designing a sequence of
excitation pulses to efficiently transfer all atoms, except for those in the dark state with
$q=0$, to the second internal state, each pulse is followed by the optical pumping process, and
repeating the sequence leads to accumulation of atoms in the dark state, i.e., to
cooling.}\label{Fig:RamanCooling} \end{figure} \end{center}

In Raman cooling \cite{raman_chu} one considers a $\Lambda$-system as shown in
Fig.~\ref{Fig:RamanCooling}, consisting of two (hyperfine) ground states $g_1$
and $g_2$ and one electronically excited state $e$. The energy of the moving
atom is given by the sum of its internal energy $\omega_i$, where
$i=g_1,g_2,e$, and its kinetic energy $q^2/2m$. The state $\ket{g_1,q}$ is
coupled to $\ket{g_2,q+(k_1+k_2)}$ via a Raman process $\hat H_{\rm LS}$, which
is far detuned from the excited state. Here, $q$ denotes the momentum of the
atom, $k_1$ and $k_2$ are the wave numbers of the two Raman laser beams and we
use units where $\hbar=1$ throughout the paper.

Cooling is achieved in a two step process. In the first step a set of Raman laser pulses is
designed that efficiently transfer atoms with high momenta in the ground state $g_1$ to $g_2$, but
do not couple atoms with zero momentum. In the second step, which is applied after each single
excitation pulse, atoms are optically pumped back to the state $g_1$ via the excited state $e$
(see Fig.~\ref{Fig:RamanCooling}). The spontaneous emission process randomizes the momentum of the
atom, which leads to a finite probability of falling into a region near the dark state, which in
this case is the state $\ket{g_1,q=0}$. Repeating the two steps leads to accumulation of atoms in
state $\ket{g_1}$ with momentum near zero, i.e., to cooling of the atom.

\subsubsection{Velocity Selective Coherent Population Trapping (VSCPT)}

\begin{center}
  \begin{figure}[htp]
    \includegraphics{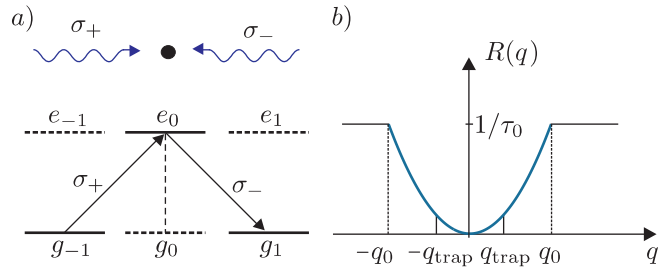}
    \caption{a) Schematic picture of the setup in VSCPT. Two circularly polarized laser beams
    incident on the atom, couple the two ground states $\ket{g_{-1}}$ and $\ket{g_1}$ via the
    excited state $\ket{e_0}$. A dark state (see text) appears due to quantum interference, where
    atoms accumulate during the cooling process as they decay there via spontaneous emissions
    from the excited state. b) Theoretical model for the excitation rate in the L\'evy statistics
    analysis. For momenta $|q|<q_0$ the excitation rate follows a power law, outside this region
    it is assumed constant.}\label{Fig:VSCPT}
\end{figure} \end{center}

In VSCPT (see Fig.~\ref{Fig:VSCPT}a)) \cite{VSCPT}, two counter-propagating circularly polarized
laser beams couple the two ground states $\ket{g_{-1}}$ and $\ket{g_1}$ via the excited state
$\ket{e_0}$. In the setup states with angular momentum $J=1$ are used in the ground and excited
state manifold, so that only the three states $\ket{g_{-1}}$, $\ket{g_1}$ and $\ket{e_0}$ are
coupled via the lasers, and spontaneous emission only leads to decay back to the states $g_{-1}$
and $g_1$ since the transition $e_0\rightarrow g_0$ is forbidden. A dark state forms due to
quantum interference and is given by $\ket{D}=(\ket{g_{-1},k_l} + \ket{g_{1},-k_l})/\sqrt{2}$,
where $k_l$ is the wave number of the two lasers. This dark state does not couple to the Raman
process and will be increasingly populated during the cooling process due to the decay of atoms in
the excited state via spontaneous emission. This leads to cooling of the atom, characterized by
two narrow peaks at $k_l$ and $-k_l$ in the final momentum distribution of the atom, as the dark
state $\ket{D}$ is a superposition of these two momentum eigenstates.

\subsection{ L\'evy statistics}\label{Section:Levy}


Analytical calculations based on L\'evy statistics \cite{Levy} have been shown to be a
very powerful and accurate tool in the context of subrecoil laser cooling. They will be
applied below to describe our optical lattice Raman cooling process. In this section we
briefly review the underlying model and some results in the context of Raman laser
cooling for the most important physical parameters, such as the temperature and the
fraction of the momentum distribution in the dark state.

In L\'evy statistics \cite{Levy} a {\em trapping region} near $q=0$, with $|q|\leq q_{\rm trap}$
and a {\em recycling region} with $|q|>q_{\rm trap}$ is defined. The parameter $q_{\rm trap}$ is
an auxiliary variable and the results for real physical quantities do not depend on it. During the
cooling process, the atom will undergo a random walk in momentum space, with $\mathcal{N}$
trapping periods of duration $\tau_i$, alternating with $\mathcal{N}$ recycling periods of
duration $\hat\tau_i$, during which the atom is in side and outside the trapping region
respectively. The total time $\Theta$ of the cooling process can be written as the sum of the
total trapping time $T_\mathcal{N}$ and the total recycling time $\hat T_\mathcal{N}$
\begin{align}
    \Theta=T_\mathcal{N}+\hat T_\mathcal{N}, \quad T_\mathcal{N}=\sum_{i=1}^\mathcal{N}
    \tau_i,\quad \hat T_\mathcal{N}=\sum_{i=1}^\mathcal{N} \hat{\tau}_i,
\end{align}
where the $\tau_i$ and the $\hat\tau_i$ are independent random variables.

The efficiency of the cooling process can be quantified by the statistics of the total
trapping and recycling times, which are determined by the excitation rate $R(q)$ of atoms
with momentum $q$ from the state $g_1$ to the state $g_2$. The rate $R(q)$ is modelled as
\begin{align}
    &R(q)=\frac{1}{\tau_0}\left|\frac{q}{q_0}\right|^\lambda, \quad |q|<q_0,\label{Rq1}\\
    &R(q)=\frac{1}{\tau_0}, \quad |q|\geq q_0,\label{Rq2}
\end{align}
as schematically plotted in Fig.~\ref{Fig:VSCPT}b). The values of $\tau_0$, $q_0$ and
$\lambda$ depend on the details of the excitation pulses and are mainly determined by the
duration, momentum transfer and shape of the last two Raman pulses. In subrecoil cooling,
typically $\lambda>1$ (especially $\lambda=2$ for the case of time square excitation
pulses in Raman laser cooling) and thus the probability distribution for the total
trapping times $T_\mathcal{N}$ is a {\it broad} distribution, i.e., the expectation
values $\EV{T_\mathcal{N}}$ and $\EV{T_\mathcal{N}^2}$ diverge. In this case, a
generalized central limit theorem (for details see \cite{Levy}) predicts a L\'evy
distribution for the probability distribution for the total trapping time.

This distribution is the starting point for a calculation of the relevant physical parameters,
like the width of the momentum distribution or the height of the central peak at $q=0$ (for
details again see \cite{Levy}), with the following results. We define the temperature in terms of
the half width $\Delta q$ at $e^{-1/2}$ of the maximum of the velocity distribution and find
\begin{align}
    \frac{1}{2}k_BT=\frac{\Delta q^2}{2m}\propto \Theta^{-2/\lambda},
\end{align} especially we find $T\propto \Theta^{-1}$ for the case of time square excitation
pulses. Similarly we derive an expression for the population density at $q=0$
and time $\Theta$, $n_0(\Theta)$, which is given by
\begin{align}
    n_0(\Theta)=\frac{\lambda^2\sin\pi/\lambda}{2\pi\gamma(1/\lambda)}
    \frac{1}{q_0}\left(\frac{\Theta}{\tau_0}
    \right)^{1/\lambda}\propto \Theta^{1/\lambda},\label{h}
\end{align}
where in this equation, $\gamma(x)$ denotes the Euler gamma function.

\section{Raman Cooling in an Optical Lattice}\label{Section:Idea}

In this section we will describe our Raman cooling scheme for atoms in an optical
lattice, therefore we will again shortly describe the idea of the cooling process and
then analyze the different parts of the setup in detail.

In our scheme cooling of the atoms is achieved in a two steps (c.f. Figs.~\ref{Fig:Scheme}a) and
\ref{Fig:Scheme}b)) and can be seen in analogy to free space Raman cooling
(Fig.~\ref{Fig:RamanCooling}). (i) Raman laser pulses are designed to excite atoms with large
quasi-momenta $|q|>0$ from the lowest band to the first excited band, whilst not exciting atoms in
the dark state (with $q=0$) (see Fig.~\ref{Fig:Scheme}a)). (ii) The decay of excited lattice atoms
goes along with the emission of a phonon (Bogoliubov excitation) into the BEC reservoir (see
\ref{Fig:Scheme}b). Assuming a BEC temperature $k_BT_b\ll \omega$ the Bogoliubov modes with an
energy corresponding to the separation of the Bloch bands are essentially in the vacuum state and
the BEC effectively acts as a $T=0$ reservoir. This is in analogy to the coupling of the atoms to
the vacuum modes of the radiation field, giving rise to spontaneous emission into a $T=0$
reservoir in the case of laser cooling. Repeating these two processes leads to cooling of the
atoms, as they accumulate in a region near the dark state with $q=0$.

Note, that in contrast to the free space version of Raman cooling, in our setup the upconversion
of energy is already performed in the excitation step. This is only possible, because in our setup
the spontaneous decay back to the lowest band is due to collisional interaction with the BEC
reservoir which can be switched off during the excitation step, e.g., via Feshbach resonances. A
direct analogue to the cooling protocol in free space Raman cooling could, however, also be
achieved with a slightly different setup: For example, one could use a spin-dependent lattice, and
perform the excitation step from the lowest Bloch band for lattice atoms $a$ to the lowest Bloch
band of a second species $a'$, followed by an optical pumping step via an excited Bloch band, as
in free space Raman cooling.

In our setup we assume a homogeneous optical lattice, i.e., no additional external
(harmonic) trapping potential for the lattice atoms. This assumption is an experimental
requirement inherent in the realization of many strongly correlated phases, and a topic
of significant current interest experimentally (see, e.g., \cite{Raizenbottomed}). Our
analysis will be performed for the ideal case of a homogeneous system, and
inhomogeneities appearing in a real experiment will limit the achievable temperatures. In
practice, we expect that advances in homogeneous traps and cooling methods will occur
somewhat iteratively, in that clean flat-bottomed traps allowing better cooling, which
will provide the opportunity to cool atoms to even lower temperatures where the remaining
imperfections become more noticeable.

\begin{center}
    \begin{figure}[htp]
        \includegraphics{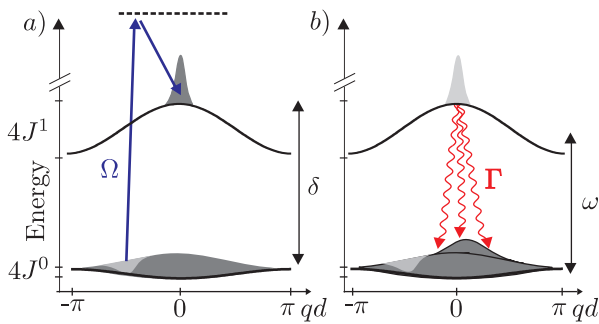}\caption{Schematic picture of one excitation
and decay step in Raman cooling in an optical lattice. a) Atoms are transferred from a region with
high quasi-momentum $|q|>0$ in the lowest Bloch band to the first excited band. b) The collisional
interaction with the BEC atoms is switched on, the resulting decay of the excited lattice atoms
leads to a randomization of the quasi-momentum. Sequences of pulses, each one followed by a decay
time $\tau_c$, efficiently excite all atoms outside a narrow region around $q=0$. Repeating the
sequence leads to accumulation of atoms in the dark state region around $q=0$, i.e., to
cooling.}\label{Fig:Scheme}
    \end{figure}
\end{center}

\subsection{Lattice atoms and laser setup}\label{Section:Ham}

The Hamiltonian $\hat H_a$ for the lattice atoms can be written as $\hat H_a=\hat H_0+\hat H_I$,
where
\begin{align}
    \hat H_0&=\int_{{\mathbbm R}^3}\ddx ~\psd{a}\left( -\frac{\nabla^2}{2m_a}+ V_a(\bx)\right)\ps{a},\label{HL0}
\end{align} describes the kinetic energy of the atoms and the optical lattice potential $V_a(\bx)=
V_{a,x}\sin^2(\pi x/d)+ V_{a,y}\sin^2(\pi y/d) + V_{a,z}\sin^2(\pi z/d)$. Here, $d=\lambda/2$ is
the lattice spacing with $\lambda$ the wavelength of the lasers generating the lattice potential,
and $V_{a,j}$ is the strength of the optical lattice potential in $j=x,y,z$ direction, $m_a$ is
the mass of the lattice atoms and $\hat \psi_a^\dagger(\bx)$ and $\hat \psi_a(\bx)$ are the field
operators for the lattice atoms, which will satisfy (anti-)commutation relations in the case of
(fermions) bosons. Onsite interactions between lattice atoms are represented by \begin{align}
  \hat H_I=\frac{g_{aa}}{2}\int_{{\mathbbm R}^3}\ddx~\psd{a}\psd{a}\ps{a}\ps{a},\label{HI0}
\end{align}
where $g_{aa}=4\pi a_{aa}/m_a$ with $a_{aa}$ the s-wave scattering length.

The interaction of the two laser beams generating the Raman transition with the
lattice atoms in the rotating wave approximation can be written as
\begin{align}
    \hat H_{\rm LS}=\int_{{\mathbbm R}^3}\ddx\Big[&\Big(\left( \frac{\Omega_1(\bx)}{2} +
    \frac{\Omega_2(\bx)}{2}\rme^{-\rmi\delta t}\right)\psd{a} \ps{e}
    \nonumber\\+ {\rm h.c.}\Big)
    &-\Delta\psd{e} \ps{e}\Big],\label{HRC0}
\end{align} where the Raman detuning $\delta=\omega_1-\omega_2$ is given by the frequency
difference of the two lasers, $\Delta$ is the detuning from the excited state, and $\Omega_1$ and
$\Omega_2$ denote the single photon Rabi frequencies of the two lasers. Both lasers are far
detuned from resonance with the transition between ground and excited state, which can thus be
adiabatically eliminated (see Appendix \ref{Appendix:adiabatelim}).

For simplicity, we will assume in the following an anisotropic lattice potential, with $V_{a,y} =
V_{a,z}\equiv V_{a,\perp} \gg V_{a,x}$ so that the atoms effectively move in one dimension along
the $x$-direction, whereas the transverse hopping is suppressed due to the large trapping
potential. Such an analysis is readily extended to higher dimensions. In the case where only the
two lowest bands play a role in the dynamics of the system, we can express the field operators in
terms of the Bloch functions for the lowest two bands in the $x$-direction, $\phi_q^{\alpha}(x)$,
where the band index $\alpha \in \{0,1\}$, and localised wavefunctions $w^0_y(y)$ and $w^0_z(z)$
representing the confinement in the transverse directions. This imposes requirements on the laser
coupling strength and detuning, and also on the interaction strength between atoms in the lattice
(see below). We obtain \begin{align}
    \ps{a}=\sum_{{\alpha=0,1}}\sum_q\phi_q^{ \alpha}(x) w^0_y(y) w^0_z(z) \A{q}{\alpha}.\label{expand}
\end{align} where $(A^\alpha_{q})^\dagger$ creates a particle with quasi-momentum $q$ in Bloch
band ${\alpha}$. The operators $(A^\alpha_{q})^\dagger$ and $(A^\alpha_{q})$ will again satisfy
(anti-)commutation relations for (fermions) bosons. Inserting this into
Eqs.~(\ref{HL0})-(\ref{HI0}), we obtain \begin{align}
      \hat H_0+\hat H_{\rm LS}&=\sum_{q,\alpha} \varepsilon_q^\alpha \left(\hat A_{q}^{\alpha}\right)^\dagger
      \hat A_{q}^\alpha+\left(\omega-\delta\right)\sum_q \left(\hat A^1_{q}\right)^\dagger \hat A^1_{q}\nonumber\\
      &+\frac{\Omega}{2}\sum_q\left[ \left(\hat A^1_{q}\right)^\dagger \hat A^0_{q-\dq}
      +{\rm h.c.}\right],\label{Ha}
\end{align}
where $\omega$ is the energy separation of the Bloch bands, and the kinetic
energy of the lattice atoms
\begin{align}
    \varepsilon^\alpha_q=-2J^\alpha \cos(qd).
\end{align} In a setup where the wave number difference of the two running wave laser
beams inducing the Raman process is parallel to the $x$-direction and of
magnitude $\dq$, we can define the effective Rabi frequency as
\begin{align}
    \Omega=\frac{\Omega_1\Omega_2}{4\Delta}\int_{-\infty}^{\infty} \dx\, {\rm exp}(-\rmi \dq
    x)w^1(x)w^0(x).
\end{align}
Here, the Wannier functions $w^\alpha(x)$ are defined as
\begin{align}
    w^\alpha(x-x_j)=\frac{1}{\sqrt{M}}\sum_q\rme^{\rmi q x_j}\phi_q^\alpha(x),
\end{align} where $x_j$ is the position of the $j$-th lattice site with $j\in \{1,..,M\}$ and $M$
is the number of lattice sites, the discrete quasi-momentum $q$ in the finite lattice $qd\in
\{-(M-1)/2,\dots (M-1)/2\}2\pi/M$. Note that we neglect contributions involving Wannier functions
in different lattice sites.

Similarly, we find that the terms describing onsite interactions between
lattice atoms can be expressed as
\begin{align}
      \hat H_I&=\frac{1}{2M}\sum_{q_1,q_2,q_3,\alpha} U^{\alpha\alpha}\Ad{q_1}{\alpha}\Ad{q_2}{\alpha}
      \A{q_3}{\alpha}\A{q_1+q_2-q_3}{\alpha}\nonumber\\
      &+\frac{2}{M}\sum_{q_1,q_2,q_3}U^{10}\Ad{q_1}{1}\Ad{q_2}{0}
      \A{q_3}{0}\A{q_1+q_2-q_3}{1},\label{HI}
\end{align}
with
\begin{align}
    U^{\alpha\alpha'}= g_{\perp}\int_{-\infty}^{\infty}\dx|w^\alpha(x)|^2 |w^{\alpha'}(x)|^2.\label{U1d}
\end{align}
Here, $g_\perp=2\omega_\perp a_{aa}$ and $\omega_\perp= \sqrt{4V_{a,\perp}
\omega_R}$ is the oscillation frequency of the transverse confinement,
$\omega_R$ is the recoil frequency. Note that this expression is valid for
$U^{\alpha \alpha'}\ll \omega$. In order to obtain the expressions for
eqs.~(\ref{HI}) and (\ref{U1d}) we have inserted eq.~(\ref{expand}) into
eq.~(\ref{HI0}) and performed the integrals over the transverse directions,
where we have approximated the lattice potential by harmonic oscillators with
frequency $\omega_\perp$.

The interaction strengths (\ref{U1d}) can be explicitly calculated if we also
approximate the Wannier functions in $x$-direction with harmonic oscillator
wave functions, and we find
\begin{align}
  U^{00}=g_\perp \sqrt{\frac{m_r}{\pi}} \left( V_{a,x}
  \omega_R\right)^{\frac{1}{4}},
\end{align} and $U^{10}=U^{00}/2$, $U^{11}=3U^{00}/4$.

In summary, the above model requires $J^\alpha, U^{\alpha,\alpha'}, \Omega\ll
\omega$, $\omega \ll \omega_\perp$.

\subsection{BEC reservoir and interaction with the lattice atoms}

The BEC-reservoir is described as a 3D homogeneous quantum gas consisting of
$N_b$ particles of mass $m_b$ in a volume $V$ by
\begin{align}
    \hat H_b=\frac{g_{bb}N_b^2}{V}+\sum_\bk E_\bk \hat b_\bk^\dagger \hat
    b_\bk,
\end{align}
where $g_{bb}=4\pi a_{bb}/m_b$, with $a_{bb}$ the scattering length for the
interaction of reservoir atoms, and the operator $\hat b_\bk^\dagger$ creates a
Bogoliubov excitation with momentum $\bk$ and energy
\begin{align}
    E_\bk=\sqrt{c^2|\bk|^2+\frac{|\bk|^4}{(2m_b)^2}},
\end{align}
in the reservoir. The sound velocity in the BEC is given by
$c=\sqrt{g_{bb}\rho_b/m_b}$, where $\rho_b=N_b/V$ is the mean condensate
density.

The interaction of the lattice atoms with the superfluid reservoir is modelled by the
density-density interaction Hamiltonian
\begin{align}
    \hat H_{\rm int}=g_{ab}\int_{{\mathbbm R}^3}\ddx\psd{a}\psd{b}\ps{b}\ps{a},
\end{align}
with the interaction strength $g_{ab}=4\pi a_{ab}/2m_r$, where $a_{ab}$ is the inter-species
s-wave scattering length and $m_r=m_am_b/(m_a+m_b)$ the reduced mass. The field operators for the
BEC can be expressed as
\begin{align}
    \ps{b}= \sqrt{\rho_b} + \delta\ps{b},
\end{align}
and
\begin{align}
    \delta\ps{b} = \frac{1}{\sqrt{V}} \sum_\bk (u_\bk \hat b_\bk
\rme^{\rmi\bk\bx} + v_\bk {\hat b}_\bk^\dagger \rme^{-\rmi \bk\bx})
\end{align}
in terms of creation and annihilation operators $\hat b_\bk^\dagger$ and $\hat b_\bk$ for a
Bogoliubov excitation with momentum $\bk=(k,k_y,k_z)$. The coefficients $u_\bk$ and $v_\bk$ can be
written as
\begin{align}
    u_\bk =\frac{1}{\sqrt{1- L_\bk^2}}, \quad v_\bk =\frac{L_\bk}{\sqrt{1-L_\bk^2}},
\end{align}
where
\begin{align}
    L_\bk=(E_\bk-\bk^2/2m_b-m_bc^2)/m_bc^2.
\end{align}
We neglect the terms proportional to $\delta\psd{b}\delta\ps{b}$ \cite{AJcooling} and using eq.~(\ref{expand}) and leaving out the constant
mean field shift we obtain
\begin{align}
  \hat H_{\rm int}=\sum_{\alpha,\alpha'}\sum_{\bk, q}\left(G_{\alpha,\alpha'}^\bk \hat b_\bk
  \left(\hat A^\alpha_{q}\right)^\dagger
   \hat A^{\alpha'}_{q-k}+{\rm h.c.} \right).\label{Hint}
\end{align}
Here, the coupling
\begin{align}
  G_{\alpha,\alpha'}^{\bk} \approx g_{ab}\left(\frac{S(\bk)\rho_b}{V}\right)^{1/2}
  \int_{-\infty}^{\infty} \dx \rme^{\rmi k x} w^\alpha(x) w^{\alpha'}(x),\label{coupling}
\end{align}
where $S(\bk)$ is the static structure factor
\begin{align}
    S(\bk)=(u_\bk + v_\bk)^2=\frac{|\bk|^2}{2m_b E_\bk}.
\end{align} In eq.~(\ref{coupling}) we have again neglected the overlap of Wannier functions in
different lattice sites and performed the integration over transverse lattice directions, where
the wave functions are again approximated with ground state harmonic oscillator wave functions and
result in factors of one, provided $m_b\omega/2m_a\omega_\perp\ll 1$, which is usually fulfilled
for our one dimensional lattice potential. The interaction Hamiltonian Eq.(\ref{Hint}) describes
scattering processes, where a Bogoliubov excitation with momentum $\bk$ in the (three dimensional)
BEC reservoir and a lattice atom with momentum $q-k$ (with $k$ the component of the momentum of
the Bogoliubov excitation along the $x$-direction) in Bloch band $\alpha'$ are annihilated
(created) and a lattice atom with momentum $q$ in Bloch band $\alpha$ is created (annihilated).

The structure factor $S(\bk)\rightarrow 1$ for energies much larger than the chemical potential
$\mu=m_b c^2=g_{bb} \rho_b$, where the corresponding Bogoliubov excitations are particle-like,
whereas for energies much less than $\mu$ the excitations are phonons and $S(\bk)\propto |\bk|
\rightarrow 0$. In our setup we will choose $4J^0\ll\mu\ll\omega$ (see Fig.~\ref{Fig:Phonons}),
which corresponds to typical experimental parameters. In this case, interband transitions will
involve absorption and emission of Bogoliubov excitations in the particle branch ($E_\bk\sim
\omega$). These will have much larger coupling strengths ($S(\bk)\sim 1$) than intraband
transitions, which involve absorption and emission of Bogoliubov excitations in the phonon branch
($E_\bk \sim J^\alpha$, $S(\bk)\propto |\bk| \rightarrow 0$) (c.f. Fig.~\ref{Fig:Phonons}).

\begin{center}
    \begin{figure}[htp]
        \includegraphics{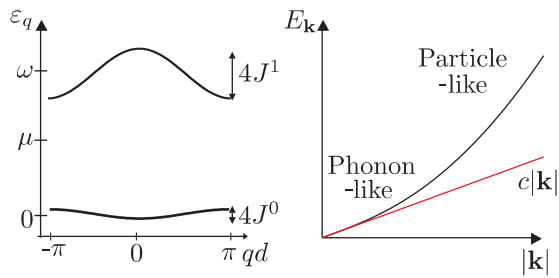}\caption{Overview over the energy scales in
the system. Left part: Energy structure in the optical lattice in the quasi-momentum picture.
Right part: The dispersion relation of the Bogoliubov excitations in the (three dimensional) BEC
reservoir. Excitations corresponding to the band separation $\omega$ are particle-like, small
excitations with energies on the order of a Bloch band width are phonon-like. The chemical
potential is larger than the width $4J^0$ of the lowest band, but less than the band separation
$\omega$.}\label{Fig:Phonons}
    \end{figure}
\end{center}

In the derivation of the master equation (section \ref{Section:ME}) we will restrict
ourselves to interband processes, i.e. to the decay from the first excited Bloch band
back to the lowest band. In the following we will comment on intraband processes, which
can lead to sympathetic heating or cooling of atoms within a Bloch band due to collisions
with Bogoliubov excitations in the BEC reservoir.

The heating/cooling process is described by the interaction Hamiltonian (\ref{Hint}) and the
corresponding rate can be estimated, e.g., with Fermi's golden rule. The heating/cooling processes
involve the scattering of a lattice atom with a single Bogoliubov excitation, and Fermi's golden
rule implies both energy and momentum conservation. We will illustrate this for the case of
sympathetic heating within a Bloch band, as this represents a possible imperfection for the
cooling process (whereas sympathetic cooling would only speed up the cooling) and the same
arguments are valid for intraband cooling. The typical case of such a heating process in our
scheme would be a scattering of an atom with momentum $q\approx 0$ and energy
$\varepsilon_{q\approx 0}^0$ to a momentum $q'$ with energy $\varepsilon_{q'}^0+c|\bk|$
(excitations within the Bloch band are sound waves), as typically most of the atoms are within a
narrow region around $q=0$ after a few iterations of the cooling process. This process is however
only allowed if energy conservation $c |\bk| =\varepsilon^0_q - \varepsilon^0_{q'}$ and momentum
conservation along the lattice axis $k=q-q'$ are fulfilled. As $c|\bk| =c| \sqrt{k^2+k_y^2+k_z^2}|
\geq c|k|$ we find that this process is forbidden if the condition
\begin{align}
    J^0 < \frac{\sqrt{\mu \omega_R m_a / (2 m_b)}}{\pi}
\end{align} is fulfilled (see Fig.~\ref{Fig:Kegel}). For the typical parameters in our setup (see
Fig.~\ref{Fig:Phonons}), where $\mu,\omega_R\gg J^0$, this condition will always be fulfilled, and
thus sympathetic heating and cooling within a Bloch band are forbidden by energy and momentum
conservation. Higher order processes arising from scattering with two or more BEC excitations will
be small \cite{AJcooling}.

\begin{center}
    \begin{figure}[htp]
        \includegraphics{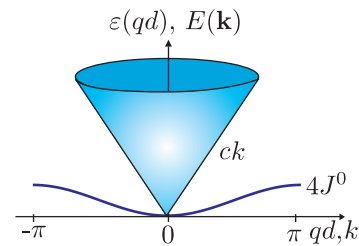}\caption{Schematic picture of the criterion for
the occurrence of sympathetic heating/cooling within the lowest Bloch band. In analogy to
the Landau criterion for superfluidity, the corresponding scattering processes are only
allowed if the Bloch band $\varepsilon(q)$ intersects the cone showing the Bogoliubov
excitation energies corresponding to the relevant momenta $\bk$.}\label{Fig:Kegel}
    \end{figure}
\end{center}

\subsection{Master equation for decay}\label{Section:ME}

The collisional interaction (\ref{Hint}) of atoms in the excited Bloch band of the optical lattice
with the BEC reservoir leads to a decay of excited lattice atoms back to the lowest band, in
analogy with spontaneous emission (see Sec.~\ref{Section:LaserCooling}). For typical BEC
temperatures $k_BT_b\ll\omega$, the Bogoliubov modes corresponding to the band separation $\omega$
will initially be in the vacuum state, and we can derive an effective $T=0$ master equation in the
Born-Markov approximation for the reduced system density operator $\hat\rho$, which describes
atoms moving in the lattice, while the BEC is treated as a reservoir of Bogoliubov excitations.
This is done in close analogy to \cite{fermiloading} and \cite{AJcooling} in the context of
lattice loading of Fermions from an external reservoir, and cooling of single atoms in a harmonic
trap immersed in a BEC, respectively. We find (see Appendix \ref{Appendix:ME})
\begin{align}\label{ME}
    \dot{\hat\rho}=\sum_k \frac{\Gamma_k}{2} \left( 2\hat c_k{\hat\rho} \hat c_k^\dagger -\hat c_k^\dagger
    \hat c_k {\hat\rho} -{\hat\rho} \hat c_k^\dagger \hat c_k\right),
\end{align} where the one dimensional momentum $k$ along the lattice axis is bounded by $|k|\leq
k_{\rm max} = \sqrt{2m_b\omega}$ due to energy conservation, and the jump operators $\hat c_k$ are
defined as \begin{align}
    \hat c_k\equiv \sum_j (\hat a^0_j)^\dag (\hat a^1_j) \rme^{-i k x_j}=   \sum_q (\hat A_{q-k}^0)^\dagger \A{q}{1},
\end{align} with the position space operators $\hat a^\alpha_i=(1/\sqrt{M}) \sum_q \exp(i q x_i)
\hat A_q^\alpha$. Note that an operator written $\hat A_{q-k}^\alpha$ should always be understood
as $\hat A_{q'}^\alpha$, where $q'$ is a quasi-momentum in the first Brillouin zone, found by
subtracting an integer multiple of the reciprocal lattice vector from $q-k$, i.e., $q'=q-k-zG,\,
z\in \mathbbm{Z}, G=2\pi/d$. Similarly, the operator $\hat c_k \equiv \hat c_{k+zG}$.

 If a single atom is
present in the excited Bloch band, the total decay rate via creation of
excitations with all possible values of $k$ is given by $\Gamma=\sum_k
\Gamma_k$. We can then define the distribution of emitted excitations, $\rmd
\Gamma/\rmd k$, which for deep lattices (where we can approximate Wannier
functions by harmonic oscillator ground states, and $\omega\gg |J^1|,J^0$) can
be written explicitly as
\begin{equation}
\frac{\rmd \Gamma}{\rmd k}\hat =\frac{L}{2\pi} \Gamma_k =
\frac{g_{ab}^2{\rho}_bm_a a_0^2k^2}{4\pi} \rme^{-a_0^2k^2/2}. \label{GammaHO}
\end{equation}
Here $a_0=\sqrt{1/m_a\omega}$ is the size of the ground state of the harmonic
oscillator in the $x$-direction and $L=Md$ is the length of the lattice along
the $x$-direction.

\begin{center}
   \begin{figure}[htp]
       \includegraphics{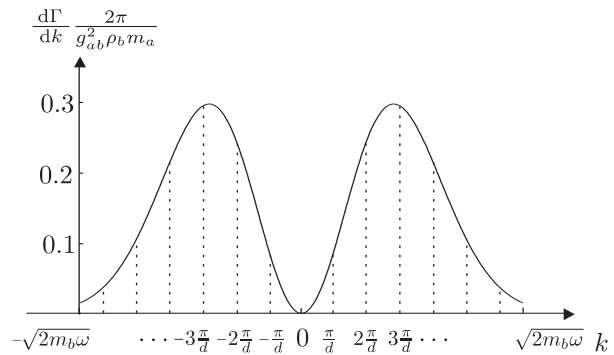}\caption{The distribution of $k$ values in decay events, $\rmd \Gamma/\rmd
       k$, plotted in dimensionless units for the case where $k_{\rm max}\gg \pi/d$. In the plot we indicate
       integer multiples of the Brillouin zone width with dashed vertical lines. In order to compute the change
       in quasi-momentum of a lattice atom $q'-q$, the values for k must always be translated back into the
       first Brillouin zone.}\label{Fig:Gammak}
   \end{figure}
\end{center}

The total decay rate can be explicitly calculated in the harmonic oscillator
approximation as
\begin{align}
  \Gamma=\frac{g_{ab}^2 \rho_bm_b}{2\pi a_{0}} \left[\sqrt{2\frac{m_b}{m_a}}
  \rme ^{-\frac{m_b}{m_a}} -\sqrt{\frac{\pi}{2}}{\rm erf} \left(\sqrt{\frac{m_b}{m_a}}
  \right) \right],\label{Gam}
\end{align}
where ${\rm erf}(x)$ denotes the error function. The value of $\Gamma$ can be
tuned by changing the scattering length, the density of the BEC reservoir or
the depth of the optical lattice.

If we compute the distribution for the change in quasi-momentum, $q'-q$ of the lattice atoms, we
must translate values for $k$ outside the first Brillouin zone, $|k|>\pi/d$ back into the first
Brillouin zone, as $q'=q-k-zG,\, z\in \mathbbm{Z}, G=2\pi/d$. This is indicated by the regions
between the dashed lines in Fig.~\ref{Fig:Gammak}. We can distinguish two interesting limits for
our parameters based on the ratio of the upper bound on $|k|$, $k_{max}=\sqrt{2 m_b \omega}$ and
the extent of the first Brillouin zone, $\pi/d=\sqrt{2 m_a \omega_R}$:

(1) $k_{\rm max}\gg \pi/d$. Here, the distribution $\rmd \Gamma/\rmd k$ extends
over $k$ values much larger than the first Brillouin zone, as depicted in
Fig.~\ref{Fig:Gammak}. When we compute the distribution of changes in
quasi-momentum for the lattice atoms, this will be approximately uniform over
the first Brillouin zone. Note that this limit also corresponds to a situation
in which the wavelength of emitted excitations, $\lambda_k=2\pi/k$ is typically
much shorter than the separation between lattice sites $d$. Thus, collective
effects of decay on different lattice sites (analogous to super-radiance and
sub-radiance in atomic decay) are suppressed.

(2) $k_{\rm max}< \pi/d$. Here, the distribution $\rmd \Gamma/\rmd k$ is cut
off before it reaches the edge of the first Brillouin zone. As a result, the
distribution of $\Delta q$ is localised at low values, peaking at the cutoff
value $k=\pm k_{\rm max}$. This can be used to target the decay to one area
within the first Brillouin zone. For example, we can choose to excite atoms to
a quasi-momentum value in the first band from which decay into the dark state
will be strongly favoured. Note that this limit also corresponds to a situation
in which the wavelength of emitted excitations, $\lambda_k > d$. Thus,
collective effects involving decay on different lattice sites occur, and this
decreased spatial resolution of the decay process corresponds to the increased
resolution that we observe in momentum space. These effects are properly
accounted for in our calculations.

\section{Single particle cooling}\label{Section:Single_Particle}

In this section we will analyze the cooling process consisting of the two steps (i) the Raman
laser excitation and (ii) the decay of excited lattice atoms. We will describe how efficient
excitation laser pulses can be designed and present the results obtained from both numerical
simulations and from analytical calculations based on L\'evy statistics. In the excitation step,
which represents the first part of our cooling protocol, we will assume that the interaction with
the BEC reservoir can be switched off (e.g., via a Feshbach resonance \cite{optfeshbach,
magnfeshbach,magnfeshbach1}), whereas the Raman coupling is switched off during the decay step.

\subsection{Designing the required laser pulses}

We define the probability $P_j(q)$ that the $j$-th pulse (with $j=0..N_p-1$ and $N_p$ the number
of pulses) excites an atom with initial quasi-momentum $q$ from the lowest band to the excited
band and require $P_j(q)=0$ for $q\approx 0$ and $P_j(q)\rightarrow 1$ for states with higher
quasi-momentum (c.f. Fig.~\ref{Fig:Scheme}a)). The probability $P_j(q)$ can be obtained by solving
the Heisenberg equations of motion for the system operators,
\begin{align}\label{HE}
    &\dot {\hat A}_q^0(t)=-\rmi\frac{\Omega_j(t)}{2}\hat A_{q+\dq_j}^1(t),\nonumber\\
    &\dot {\hat A}_{q+\dq_j}^1(t)=\rmi\delta_{q+\dq_j}\hat A_{q+\dq_j}^1(t)-
    \rmi\frac{\Omega_j(t)}{2}\hat A_{q}^0(t),
\end{align}
where the effective detuning $\delta_{q+\dq_j}\equiv \omega+ \varepsilon_{q+\dq_j}^1 - \varepsilon_q^0- \delta$.

These equations are valid in the subspace of a single lattice atom, where the interaction
Hamiltonian $\hat H_I=0$, and can be analytically solved for two simple cases. In the case of weak
excitation, $\int\Omega_j(t)\dt\ll 1$, we find probability
\begin{align}
    P_j(q)=\left|\half\int_{-\infty}^{\infty}\dt~\Omega_j(t)\rme^{\-\rmi \delta_{q+\dq_j} t}\right|^2
\end{align} in terms of the Fourier transform of the Raman Rabi frequency $\Omega_j(t)$. In the
case of a time square pulse ($\Omega_j(t)=\Omega_j$ for $0 \leq t\leq \tau_j$ and $\Omega_j(t)=0$
otherwise), we have
\begin{align}
  P_j(q)= \frac{\Omega_j^2}{(\delta_{q+\dq_{\!j}}^2+\Omega_j^2)}
  \sin^2\left(\sqrt{\delta_{q+\dq_{\!j}}^2+\Omega_j^2}\tau_j/2\right).
\end{align}

\begin{center}
    \begin{figure}[htp]
        \includegraphics{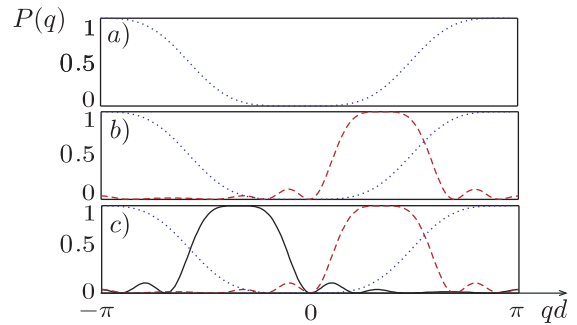}\caption{A typical efficient pulse sequence
consisting of three time square pulses. a) A short pulse excites atoms with large
quasi-momentum at the edges of the Brillouin zone (dotted line), dashed line in b) and solid line in c), longer pulses excite atoms with quasi-momentum closer to zero, always keeping $P(q=0)=0$. Parameters used: $\Omega=(27.4,8.4,8.4)J^0$, $\dq d=(0.31,2.12,-2.12)$,
$(\delta-\omega)=-(28.2,25.2,25.3)J^0$}\label{Fig:Pulses}
    \end{figure}
\end{center}

The goal of the excitation step is to design efficient laser pulses, which excite atoms with large
quasi-momentum $|q|>0$ but do not couple the atoms in the dark state $q=0$. We would also like to
do this on a fast timescale and therefore require a $\pi$-pulse, i.e., $\tau_j=\pi/\Omega_j$ in
the case of a time square pulse, for the resonant transition and adjust the parameters to always
keep the first node of the sinc-function at $q=0$. Such a pulse sequence consisting of $N_p=3$
pulses, is shown in Fig.~\ref{Fig:Pulses}. We start with an intense laser pulse to excite atoms
with momentum $qd\sim \pi$ around the edges of the Brillouin zone, and then move the resonance
closer to $q=0$ by adjusting the Raman detuning $\delta_j$ and momentum kick $\dq_j$ and at the
same time reducing the intensity of the laser beams. To be able to resolve the band structure with
our Raman pulses we always have to fulfill $\Omega\ll 8|J^1|$ and consequently $\tau\gg
\pi/8|J^1|$. Note, that the relevant energy scale is the hopping $|J^1|$ in the upper band, which
is typically an order of magnitude larger than the hopping $J^0$ in the lower band for lattice
strengths $V_{a,x}\sim 10\omega_R$. The parameters here have also been carefully chosen to avoid
unwanted excitation to higher bands.

\subsection{Results}

In this section we will quantitatively analyze the cooling process, computing
the final temperature and cooling timescales. We make use of both numerical and
analytical methods, and compare the results we obtain in each case.

In the numerical analysis we simulate the time evolution of the system density
operator using a monte carlo wave function method \cite{QN,QKT}. In the
simulations we start with an initial mixed state according to a thermal
distribution of atoms in the lowest Bloch band with a typical temperature
$4J^0\ll k_BT\ll\omega$. To obtain good approximations of the system density
operator at all times we evolve the state according to the master equation
(\ref{ME}) and typically take the statistical average over $\sim 10^5$
trajectories in the simulations, for a one-dimensional optical lattice, with
$M=101$ lattice sites and two bands.

The analytical calculations make use of L\'evy statistics, similar to the corresponding
calculations for freespace subrecoil laser cooling (see section \ref{Section:Levy}). We define the temperature of the atom in terms of the width of the quasi-momentum distribution, and find
\begin{align}
  k_BT=2J^0(\Delta\! q d)^2,\label{Tdef}
\end{align} with $\Delta\! q$ denoting the half width of the momentum distribution at
$\rme^{-1/2}$ of the maximum. This is in close analogy to free space Raman cooling \cite{Levy},
with the mass in free space replaced with the effective mass in the optical lattice, and the
quasi-momentum now playing the role of momentum in free space. We again find $T\propto
\Theta^{-2/\lambda}$ (see section \ref{Section:Levy}, and consequently time square excitation
pulses again lead to efficient cooling, as $\lambda=2$ for time square pulses also in the presence
of the optical lattice. Similarly we again find $n_0(\Theta)\propto\Theta^{1/\lambda}$.

In Fig.~\ref{Fig:singleresults} we compare the analytical and the numerical results for the
temperature and the fraction of atoms in the dark state. In the simulations, repeated application
of the $N_p$ excitation and decay steps leads to the development of a sharp peak in the momentum
distribution $n^0(q)$ in the lowest Bloch band already after a few iterations, as can be seen in
Fig.~\ref{Fig:singleresults} a).

\begin{figure}[htp]
\begin{center}
    \includegraphics{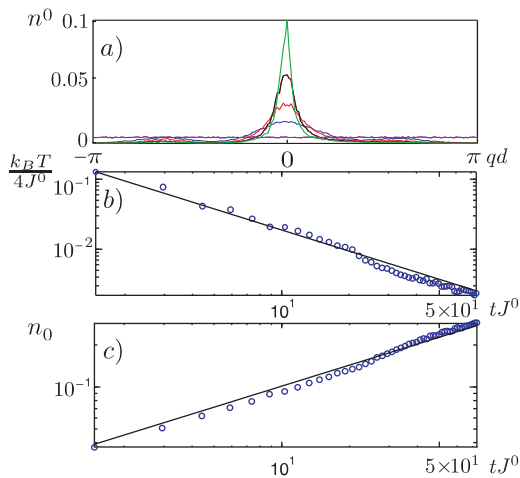}\caption{a) Development of the narrow
momentum peak after $0$, $1$, $3$, $5$ and $10$ iterations of the pulse sequence shown in
Fig.~\ref{Fig:Pulses}, for $M=101$ lattice sites. b) Temperature in units of $4J^0$ against time
in units of $1/J^0$. c) Fraction of the atoms in the dark state with $q=0$ against time in units
of $J^0$. In both plots, circles show numerical results obtained from quantum monte carlo wave
function simulations, the solid line shows analytic results obtained from L\'evy statistics.
Parameters as in Fig.~\ref{Fig:Pulses}, again for $M=101$ lattice sites.}\label{Fig:singleresults}
\end{center}
\end{figure}

In Fig.~\ref{Fig:singleresults}b) and c) we plot the temperature of the lattice atom, as defined
in eq.~(\ref{Tdef}) and the height of the peak at $q=0$ against time in units of $J^0$, where
circles denote numerical results from the quantum trajectory simulations, and solid lines are the
analytical results from L\'evy statistics. In both cases we find excellent agreement of the
numerical results with the analytical predictions. Typical temperatures of $k_BT/4J^0\sim 2\times
10^{-3}$ and a typical fraction on the order of a few tens of percent in the central peak can be
reached in $t_fJ^0\sim 50$ for the pulse sequence shown in Fig.~\ref{Fig:Pulses}. In the
simulations we have furthermore used $\Gamma=1\omega_R$, which can be obtained from eq.(\ref{Gam})
for $\rho_b=5\times 10^{14}$cm$^{-3}$ and $a_s=350a_b$ (e.g., via a Feshbach resonance) for $^{\rm
87}$Rb in the lattice and $^{\rm 23}$Na in the reservoir, $a_b$ is the Bohr radius. Note, that for
the parameters we use (see caption in Fig.~\ref{Fig:Pulses}), the cooling timescale is mainly
determined by the duration of the last two excitation pulses. The cooling timescale is typically
much faster than the excitation pulses, and thus smaller values of $\Gamma$ does not significantly
slow down the cooling process, unless the cooling timescale becomes comparable to the duration of
the excitation pulses.

\begin{figure}
  \begin{center}
    \includegraphics{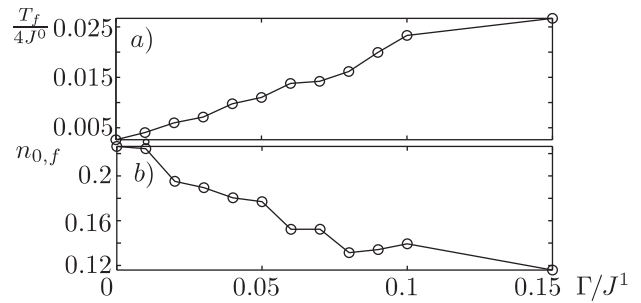}\caption{a) Temperature and b) dark state population after a time
    $t=70J^0$ for the parameters as in Fig.~\ref{Fig:Pulses} as a function of the residual decay
    $\Gamma$ obtained from numerical simulations as described in the text. }\label{Fig:finiteGamma}
  \end{center}
\end{figure}

In our cooling scheme we assume that the interaction of lattice atoms with the reservoir can be
switched off during the excitation process. In a real experiment this can be done, e.g., via
optical or magnetic Feshbach resonances \cite{optfeshbach,magnfeshbach,magnfeshbach1}, eventual
residual finite interactions (e.g., due to magnetic field fluctuations for magnetic Feshbach
resonances) can lead to a change of the excitation profile and thus represent a possible source of
imperfection. From numerical simulations, however, we find that for $\Gamma\ll J^1$, the final
temperatures (see Fig.~\ref{Fig:finiteGamma}a)) and the fraction of atoms in the dark state (see
Fig.~\ref{Fig:finiteGamma}b)) do not change significantly. From eq.~(\ref{Gam}) we find that
$\Gamma \sim 10^{-3}J^1$ can be achieved for $\rho_b=5\times 10^{14}$cm$^{-3}$ and $a_s=20a_b$ for
$^{\rm 87}$Rb in the lattice and $^{\rm 23}$Na in the reservoir.

\section{Many particle cooling}\label{Section:NParticles}

In this section we will adapt the cooling scheme to many quantum degenerate
bosons or fermions. We will assume that no interactions between the lattice
atoms are present during the cooling process. In the case of bosons this means
that the interaction Hamiltonian $H_I=0$ (see eq.~(\ref{HI})), which can be
achieved experimentally, e.g., via an optical or magnetic Feshbach resonance
(\cite{optfeshbach,magnfeshbach,magnfeshbach1}). We consider a single species of Fermions,
and thus s-wave interactions are forbidden due to Pauli blocking.

The cooling process again follows the protocol as in the case of a single lattice atom: (i) In the
excitation step Raman laser pulses are designed to excite atoms outside the dark state region to
the first excited Bloch band, the coupling of the lattice atoms to the reservoir atoms are
switched off during this step. (ii) In the decay step the dissipative coupling to the BEC
reservoir randomizes the quasi-momentum of lattice atoms decaying to the lowest band and
consequently to a finite probability of falling into the dark state region.

For non-interacting lattice atoms, the dynamics of the cooling process is again
described by the master equation (\ref{ME}). However, exact numerical
simulations of the master equation are impractical, as the discretisation of
the momentum space grid must be very fine to make possible calculation of the
low final temperatures. Therefore, we perform the analysis of the cooling
process by numerical simulations of a quantum Boltzman master equation (QBME)
\cite{QKT,QKTII}. The QBME is one of the simplest versions of the more general
quantum kinetic master equation (QKME) \cite{QKT}, which represents a fully
quantum mechanical kinetic theory for the time evolution of the system density
operator, and was originally developed to analyse formation of Bose-Einstein
condensates in atomic gases. The QBME is an equation for the diagonal elements
$w_\bm\equiv \bra{\bm}\rho\ket{\bm}$ of the reduced system density operator,
which describes the time evolution of the system in terms of classical
configurations $w_\bm$ of atoms occupying momentum states $\bm= [\{ m_q^0\}_q,
\{m_q^1\}_q]$ in the two Bloch bands. Here $m_q^\alpha$ denotes the occupation
of momentum state $q$ in Bloch band $\alpha$. In addition to the Born
approximation and the Markov approximation made in the derivation of the master
equation, the QBME neglects the off-diagonal coherences between different
classical configurations contained in the QKME.

In our numerical simulations we require a full QBME only for the decay step
(ii), as for non-interacting atoms the excitation step (i) can be exactly
calculated from the Heisenberg equations (\ref{HE}), i.e., from the excitation
probability $P_j(q)$ as \begin{align}
    w_{m_q^1}=P_j(q-\dq_j) w_{m_{q-\dq_j}^0}.\label{QBMEex}
\end{align} The QBME for the second step (ii), the decay of excited lattice atoms back to the
lowest Bloch band can be written as (see Appendix \ref{Appendix:QBME})
\begin{align}
    \dot{w}_{\bm}=&\sum_{k,q}\Gamma_k\left[ \rmm_{q-k}^0(1\pm\rmm_{q}^1)w_{\bm'}
    - \rmm_{q}^1 (1\pm\rmm_{q-k}^0)w_{\bm}\right],\label{QBME}
\end{align}
where $\bm'=\bm-\be_{q-k,q}$ and
\begin{align}
    \be_{q-k,q}=[0,...0,\stackrel{q-k}{1},0,...0,\stackrel{q}{-1},0...0],
\end{align}
the upper (lower) signs are for bosons (fermions).

Finally, we want to remark that the coherences which are neglected in the
description of the decay step in terms of a QBME could, in principle, even be
destroyed in a real experiment. For example, this could be done, by modulating
the lattice depth and thereby randomizing the off diagonal elements, similar to
the twirl in state purification protocols \cite{twirl}.

In the following we will present the results obtained from our numerical
simulations first for the case of non-interacting bosons, then we will describe
how finite interactions affect the excitation steps and finally we will present
the results for spin polarized fermions.

\subsection{Results}

\subsubsection{Non-interacting bosons}

For $N$ non-interacting bosons, the $T=0$ ground state is the fully occupied
$q=0$ momentum state, as in the case of a single atom. As a consequence we can
use the same excitation pulse sequences as for a single atom (c.f.
Fig.~\ref{Fig:Pulses}). In Fig.~\ref{Fig:Nbosons}a) we plot temperature against
time which we obtain from numerical simulations \cite{QKT} of the QBME
eq.~(\ref{QBME}). Temperature here is calculated by a Gaussian fit to the
momentum distribution, excluding the central peak. Cooling to similar
temperatures as in the single particle case can be obtained on shorter
timescales for many atoms. This is due to the bosonic enhancement factor, which
appears as the factor $(1+m_{q-k}^0)$ in the QBME (\ref{QBME}). In
Fig.~\ref{Fig:Nbosons}b) we show the increase of the central peak against time.

\begin{center}
\begin{figure}[htp]
 \includegraphics{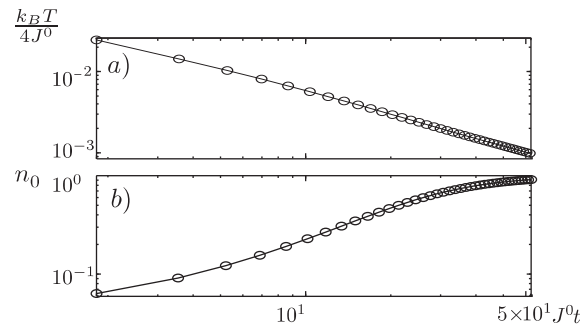}\caption{Numerical simulation of the QBME for
   $N=51$ bosons in $M=101$ lattice sites. a) Temperature in units of
$4J^0$ against time in units of $1/J^0$. b) Increase of the fraction of atoms
in the dark state region $qd\in [-0.06,0.06]$ against time in units of $1/J^0$.
Parameters used as in Fig.~\ref{Fig:Pulses}.}\label{Fig:Nbosons} \end{figure}
\end{center}

\subsubsection{Excitation profile for bosons with interactions}

Until now we we have assumed that the interactions between the Bosons $a$ is
negligible. In this section we investigate how small finite interactions will
alter the excitation profile, and give approximate values for interaction
strengths that can be safely neglected. Here, we must compute the time
evolution of the many-body system during the excitation step, which is
described by the full two-band Bose Hubbard Hamiltonian for interacting atoms
given in section~\ref{Section:Ham} as $\hat{H}_0 + \hat{H}_{LS} + \hat{H}_I$.

This is possible in 1D at temperature $T=0$ using time-dependent DMRG methods
\cite{DMRGVidal,DMRGWhite,DMRGAJ}. These numerical methods allow us to compute
the time evolution of a 1D many-body system where the Hilbert space can be
expressed as the product of local Hilbert spaces of dimension $\mathbbm{d}$
forming a 1D chain. In these methods, the state of the system is written as a
truncated Matrix Product state representation \cite{DMRGVidal}, in which $\chi$
states are retained in each Schmidt decomposition of the system. The dynamical
evolution for 1D systems with sizes similar those seen in experiments can then
be computed, starting both from the ground state and from weakly excited
states.

We simulate the time evolution in this way for the duration of one coherent
Raman pulse. We consider the situation with all the interactions equal, i.e.
$U^{00}=U^{01}=U^{11}$. To minimize the influence of box boundary conditions we
used $M=41$ sites and an initial Fock state of the form
$\ket{0\cdots0111110\cdots0}$ with $N=5$ atoms located at the centre of the
system in the lower band. Since Fock states have a flat equally occupied
momentum distribution they allow the excitation profile to be determined by
examining the final momentum distribution. A $\chi=50$ and an occupancy cut-off
of $n^{0}_{\textrm{max}}\leq 4$ and $n^{1}_{\textrm{max}}\leq 2$ atoms per site
for the lower and upper band respectively (equivalent to $\mathbbm{d}=15$) was
found to be sufficient.

In Fig.~\ref{fig:profile}a) the momentum distribution for the lowest band $n^0(q)$ after the pulse
has been applied is displayed for a sequence of interaction strengths $U^{\alpha\alpha'}/J^0$. For
the parameters chosen when $U^{\alpha\alpha'}=0$ the initially flat momentum distribution has a
hole carved out at $q = 4\Delta q$, where $\Delta q = 2\pi/Md$. This demonstrates the selective
excitation of this momentum and crucially the dark state at $q=0$ is left unchanged. As
interactions are switched on the pulse begins to distort the final momentum distribution from this
characteristic shape with a peak in the population of momenta either side of $q=4\Delta q$ as well
as an increase in the population remaining at $q=4\Delta q$ itself. This latter quantity provides
a useful measure of the shape of the excitation profile and is plotted in Fig.~\ref{fig:profile}b)
and displays a linear increase only after $U^{\alpha\alpha'}/J^0 > 0.2$. The most relevant
quantity for the cooling scheme is $n^0(0)$, the population at $q=0$, which is seen to increase
linearly for small interaction strengths in Fig.~\ref{fig:profile}c). In total this reveals that
for $U^{\alpha\alpha'} \ll 1/\tau \ll |J^1|$ the excitation profile remains close to that assumed
for $U^{\alpha\alpha'}=0$. If we include the additional constraint that $U^{\alpha\alpha'} \ll
|J^0|$, which ensures that interactions do not substantially alter the ground state, the conclusions
for the excitation pulse used earlier should not change.


\begin{center}
\begin{figure}[htp]
\includegraphics{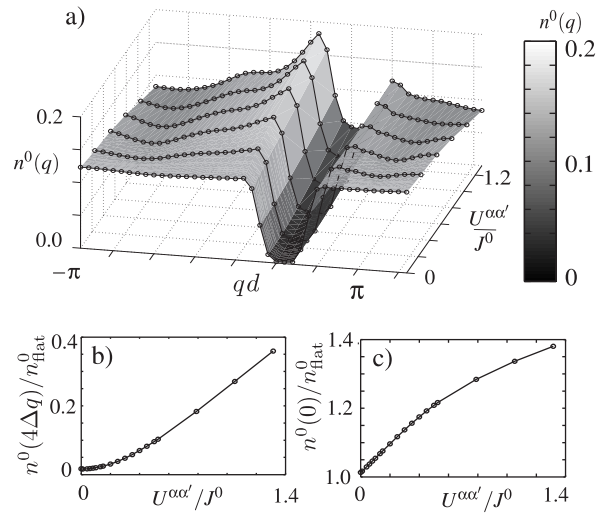}
\caption{a) The momentum distribution in the lowest Bloch band obtained after a single excitation
pulse applied is to an initial Fock state of $N=5$ atoms in $M=41$ lattice sites for a variety of
interaction strengths $U^{\alpha\alpha'}/J^0 \in [0,1.3]$. The parameters used where $\delta q d =
1.23$, $\omega = 1.05 J^0$, $(\delta - \omega) = 27.9 J^0$, $V_0 = 10\omega_R$, $\omega_R = 2\pi
\times 3.8$kHz. b) The occupancy at momentum $q=4\Delta q$ and c) at $q=0$ against the interaction
strength as a fraction of the flat occupation $n^0_{\textrm{flat}}=N/M$.} \label{fig:profile}
\end{figure}
\end{center}

\subsubsection{Fermions}

In contrast to bosons, in the case of $N$ non-interacting (spin-polarized)
fermions the $T=0$ ground state is characterized by the $T=0$ Fermi
distribution, i.e., a step function $n^0(q)=\Theta(q-q_F)$ for the momentum
distribution in the lowest band, where the Fermi momentum $q_F=\pi(N-1)/Md$.
The excitation pulse sequence thus has to be changed in order to create a dark
state region for atoms with momenta $|q|\leq q_F$. As a consequence the use of
time square pulses is no longer advantageous, due to the large sidelobes they
create in the excitation probability. We thus use a sequence of Blackman pulses
\cite{raman_chu} where these sidelobes are suppressed, as shown in
Fig.~\ref{Fig:Blackman}, where we compare typical excitation probabilities for
a Blackman pulse and a time square pulse. The sequence of excitation pulses is
now designed as in the previous cases, however the excitation probability can
only be calculated numerically and the $\pi$-pulse condition changes to
$\tau=\pi/0.42\Omega$ for Blackman pulses \cite{Levy}.

\begin{center}
    \begin{figure}[htp]
        \includegraphics{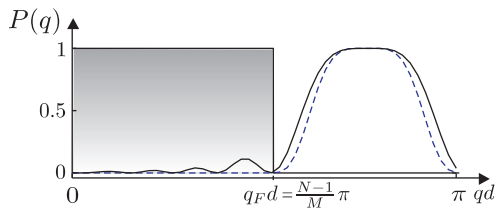}\caption{Comparison of the excitation
probability for a time square (solid line) and a Blackman pulse (dashed line), shaded area depicts the filled Fermi sea. Parameters used: $\Omega=4.21J^0$, $\dq d=0.84$, $(\delta-\omega)=-27.6$, $M=101$, $N=51$.}\label{Fig:Blackman}
    \end{figure}
\end{center}

In Fig.~\ref{Fig:Nfermions}a) we plot the momentum distribution (again obtained from monte carlo
simulations of the QBME) after $j=0,1,2,20$ cooling cycles and find that the expected shape close
to the expected $T=0$ Fermi distribution appears after a few cooling cycles. Temperature is now
obtained by fitting a Fermi distribution to occupation of the momentum states in the lowest Bloch
band. In Fig.~\ref{Fig:Nfermions}b) we plot temperature against time in units of $J^0$ and find
that temperatures $k_BT/4J^0\sim 10^{-2}$ can be obtained in $tJ^0\sim 500$ for the parameters
used (see caption of Fig.~\ref{Fig:Nfermions}). The timescales for many fermions are slower than
those of bosons due to Pauli blocking, which increasingly slows down the decay into an
increasingly filled Fermi sea.

\begin{center}
    \begin{figure}[htp]
        \includegraphics{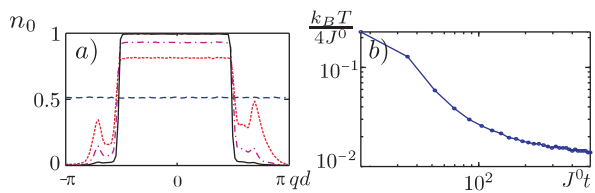}\caption{Numerical simulation of the QBME
for $N=51$ fermions in $M=101$ lattice sites. a) Development of a sharp Fermi distribution after
$0$ (dashed line), $1$ (dash-dotted line), $5$ (dotted line) and $20$ (solid line) cooling cycles. b) Temperature in units of $4J^0$ against time in units of $1/J^0$.
Parameters used: $\Omega=(11.58,11.58,1.1,1.1)J^0$, ($\dq d= 0.31,-0.31,1.2,-1.2)$, $(\delta-
\omega)=-(28.19,28.19,26.89,26.89)J^0$.}\label{Fig:Nfermions}
    \end{figure}
\end{center}

\section{Realizing cold strongly-interacting gases}\label{Section:ramping}

In this section we investigate how strongly correlated regimes can be realized
where the cooling scheme cannot be applied directly. This can be done by
decoupling the optical lattice from the reservoir $b$ and adiabatically ramping
up the interaction strength. In this case our attention is restricted to the
evolution governing by $\hat{H}_0+\hat{H}_I$ with no Raman coupling and
$\alpha=0$, and so describes the Bose-Hubbard model of the lowest band only. We
analyse this by again making use of time-dependent DMRG methods
\cite{DMRGVidal,DMRGWhite,DMRGAJ}.

\subsection{Ramping the ground state}\label{Section:groundstate}

In principle adiabaticity requires an infinitely slow ramping. Here we
determine a finite timescale in which near-adiabatic ramping can be achieved.
We consider a 1D lattice of $M=21$ sites containing $N=10$ atoms initially
prepared in the ground state with $U^{00}=0$ and then raise the interaction
strength over a time $\tau_r$ according to
\begin{equation}
\label{ramping} U^{00}(t) =
U_{\textrm{max}}\left\{1-\left[1+\exp\left(\frac{t-\tau_r/2}{\tau_r/w}\right)\right]^{-1}\right\}+\mathcal{C}.
\end{equation}
The constant $\mathcal{C}$ is fixed so as to make $U^{00}(t=0) = 0$ and we
choose the parameter $w=18\,\tau_r$ so that $U^{00}(t=\tau_r) \simeq
U_{\textrm{max}}$. As this simulation begins from a non-interacting limit we
took the occupancy cut-off to be $n^{0}_{\textrm{max}} \leq 6$ atoms per site
(equivalent to $\mathbbm{d}=7$). We found that for the computation of
groundstates retaining $\chi=50$ states in the matrix product decomposition was
sufficient for this system, however, to accurately describe the dynamical
ramping a $\chi=100$ was required. For the ramping itself we took
$U_{\textrm{max}}=20\,J^{0}$ and so approached the hard-core Bose lattice gas
in 1D (Tonks gas).

To quantify the achievement of near-adiabatic ramping we computed the energy deposited into the
system $\mathcal{E}(\tau_r)=[E(\tau_r)-E_0]/NJ^0$ and the many-body overlap $\mathcal{F}(\tau_r)
=|\braket{\psi(\tau_r)}{\psi_0}|$ as a function of the ramping time $\tau_r$ of the final ramped
state $\ket{\psi(\tau_r)}$ with energy $E(\tau_r)$ and the groundstate $\ket{\psi_0}$ with energy
$E_0$ for $U^{00}(\tau_r)$. Note that we express all energy differences $\mathcal{E}$ as a
fraction of $NJ^0$. This is a useful energy scale since it is (up to a constant prefactor of order
1) the maximum energy difference which $N$ non-interacting atoms inside the lowest band could
have. Both $\mathcal{E}(\tau_r)$ and $F(\tau_r)$ are plotted together in Fig.~\ref{fig:ramping}a).
We observe that for $\tau_r > 10/J^0$ there is an exponential decrease in $\mathcal{E}(\tau_r)$
with $\tau_r$, and thus for $\tau_r$ of order $U^{00}(\tau_r)/J^0$ there is negligible heating
within the Bloch band. Additionally $F(\tau_r)$ can be seen to rapidly approach unity on the same
timescale rigourously verifying that this final state is converging to the groundstate in the
strongly correlated regime. We note, however, that since these calculations are based on a single
system size they do not address the issue of scaling with $M$. Despite this we expect the results
presented to be applicable to systems which are currently realized in experiments since their size
is typically of the same order as considered here.



\begin{center}
\begin{figure}[htp]
\includegraphics{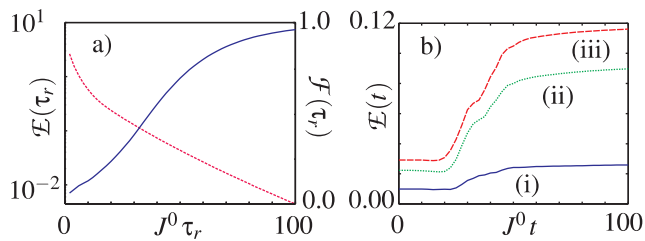}
\caption{a) The normalized energy difference $\mathcal{E}(\tau_r)$ (left-axis and dashed line) and
the many-body overlap $\mathcal{F}(\tau_r)$ (right-axis and solid line) as a function of $\tau_r$
for a ramping of the $U^{00}(t=0) = 0$ groundstate with $N = 10$, $M = 21$ to $U^{00}(\tau_r) =
20\,J^{0}$ according to the profile given in Eq. \ref{ramping}. b) The normalized energy
difference $\mathcal{E}(t)$ with the instantaneous ground state at time $t$ of the ramping with
$\tau_r=100/J^0$ for initial states generated with $J^0\,t_{\textrm{ex}}$ equal to (i) $1$, (ii)
$5$ and (iii) $9$ respectively. These weakly excited SF states are characterized by
$\mathcal{E}=\{0.01,0.02,0.03\}$, $\mathcal{D}=\{1.2\%,5.7\%,7.6\%\}$ and
$\mathcal{F}=\{0.95,0.69,0.54\}$ respectively for (i), (ii) and (iii). All three initial states
had $\Delta\mathcal{E}\approx 0.06$.}\label{fig:ramping}
\end{figure}
\end{center}

\subsection{Ramping weakly excited states}\label{Section:excited}

In the previous section the degree of excitation induced by the ramping process on the ground
state was quantified. Here we confirm that the near-adiabatic timescale determined for the ground
state also applies to good approximation to low-lying excited states. This is done by performing
the ramping with a weakly excited initial state and demonstrating that the resulting final state
remains weakly excited in the strongly interacting regime. Specifically, we generate an excited
state by evolving the $U^{00}=0$ ground state for a time $t_{\textrm{ex}}$ in the presence of a
spatially homogeneous on-site interaction which is varying randomly in time in the range
$U^{00}(t) \in [0,\half J^0]$. With this method we constructed three excited initial states using
$J^0\,t_{\textrm{ex}}$ equal to (i) 1, (ii) 5 and (iii) 9. We characterize these states by their
normalized energy difference $\mathcal{E}$, many-body overlap $\mathcal{F}$ with the ground state,
quantum depletion $\mathcal{D}$ (equal to $N$ minus the largest eigenvalue of the single-particle
density matrix), and energy spread defined by
$(\Delta\mathcal{E}/NJ^0)^2=\av{\hat{H}^2}-\av{\hat{H}}^2$ (see caption of Fig.~\ref{fig:ramping}
for specific numbers). Despite this being classical noise which coherently excites the ground
state it does produce excited states with features similar to those found in experiments. In
particular the characteristics for state (i) may be representative of a weakly excited SF state
resulting from the cooling scheme outlined.

The three initial states were then ramped in an identical way to section~\ref{Section:groundstate}
using $\tau_r = 100/J^0$. In Fig.~\ref{fig:ramping}b) the evolution of energy difference
$\mathcal{E}(t)$ as a function of the time $t$ during a ramp is shown for each of them. From the
two most excited of these states, (ii) and (iii), $\mathcal{E}(t=\tau_r)$ is seen to level-off at
around 4 times their initial values giving on the order of $10\%$ heating within the Bloch band,
and have $\Delta\mathcal{E}(t=\tau_r)\simeq 0.17$, while the many-body overlaps of their final
ramped states with the strongly-correlated ground state reduce to $\mathcal{F}=0.55$ and
$\mathcal{F}=0.41$ respectively. This indicates that the ramping is not entirely adiabatic for
these states. For the least excited initial state (i) we find that both $\mathcal{E}(t=\tau_r)$
and $\Delta\mathcal{E}(t=\tau_r)$ have approximately doubled, but crucially the heating
$\mathcal{E}(t=\tau_r)$ is still less than the $4\%$. Finally, for (i) the many-body overlap
$\mathcal{F}=0.89$ suffers a smaller reduction and importantly remains sizable sizable showing
that the final ramped state is still weakly excited in the strongly-correlated regime.

\section{Spatial sideband cooling in a harmonic trap}\label{Section:compacting}

Up until now we have discussed the implementation of a Raman cooling scheme in
an optical lattice with a flat external potential, based on the coupling of
atoms in a lattice to a reservoir gas. As an additional remark, we describe in
this section how these ideas can be applied in a different way to cool atoms in
a harmonic trap to sites of lower energy, in analogy with sideband cooling to
lower motional states in an ion trap. \cite{LCstandingwave}. This example
serves to strengthen the formal analogies between open quantum systems
encountered in quantum optics, and systems of atoms in optical lattices coupled
to a phonon bath. It can also be applied in the context of preparation of a
quantum register with Fermions \cite{fermiloading,calarcoloading}, in this case
compacting the cloud of atoms as they collect in the centre of the trap, at the
sites of lower energy (see Fig.~\ref{Fig:compact}a). Here we outline how single
atoms can be cooled to sites of lower energy, and then briefly discuss the
application to many fermions.

We consider the same setup discussed in section \ref{Section:Introduction},
with an identical Hamiltonian, Eq. (\ref{Ha}), except that here we impose an
additional external trapping potential (normally a harmonic trap), $\hat
H_E=\sum_i \epsilon_i \hat{n}_i$, where $\hat{n}_i$ is the number operator on
site $i$. In this scheme, Raman coupling of atoms from the ground motional
state in each lattice site to the excited motional state is not pulsed, but
continuously switched on. The interaction with the external reservoir gas $b$,
giving rise to the master equation (\ref{ME}) is also identical, and is also
continuously switched on, giving rise to onsite decay. We assume here that the
lattice depth is large, so that we are in a limit where collective processes
(analogous to superradiance and subradiance in phonon emission) can be
neglected (see section \ref{Section:ME}).

\begin{center}
    \begin{figure}[htp]
        \includegraphics{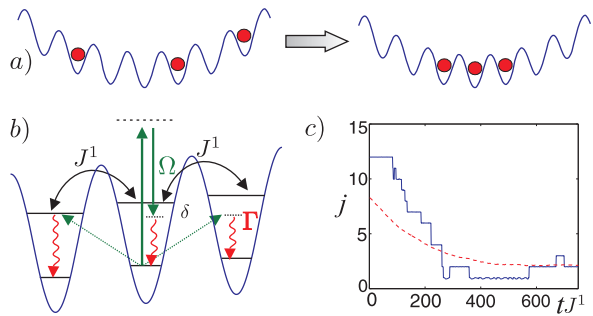}\caption{(a) Techniques involving spontentous creation of excitions in the reservoir can
        also be used to cool Fermions in an external trapping potential
        to sites of lower energy, compacting them so that in the centre of the trap we observe unit occupancy.
        (b) Atoms are excited to the first excited motional level in each well, by a Raman transition (with two-photon
        Rabi frequency $\Omega$) detuned to energies below resonance with the excited state. Atoms can then tunnel to
        neighbouring sites with tunnelling rate $J^{1}$, and can decay back to the ground motional state by interaction
        with the external reservoir gas. The dashed arrows show the effective result of the two-step excitation and tunnelling
        process, and we see that transitions to sites of lower energy are favoured because these are closer to resonance.
        (c) Results for a single particle being cooled in this manner: Plot of the mean position of the atom in a single quantum trajectory as a function of time, beginning with an atom displaced $12$ sites from the centre of a
        harmonic potential, $\epsilon_j=0.25 j^2 \,J^1$ (solid line), and root-mean-square displacement of the atom from the centre of
        the trap, averaged over an initial uniform distribution between sites $j=-15$ and $j=15$ (dashed line). Parameters used: $\Omega=J^1$, $\Gamma= 2 J^1$, $\delta= -4 J^1$ }\label{Fig:compact}
    \end{figure}
\end{center}

The basic concept of this process is shown in Fig.~\ref{Fig:compact}b. We
choose the detuning of the Raman excitation so that it is below resonance with
the excited motional level. If the atom subsequently tunnels (with amplitude
$J^1$) to a neighbouring lattice site, and then undergoes a decay, it can be
transferred to neighbouring lattice sites. (Note that tunnelling in the lowest
band $J^0 \ll J^1$, and is further suppressed by the potential offset between
the lattice sites, due to which tunnelling in the lowest band is no longer a
resonant process.) Transfer to sites of lower energy is favoured, however,
because of our choice of detuning for the Raman excitation. In
Fig.~\ref{Fig:compact}b, dashed arrows indicate an effective two-step process
including the Raman excitation and tunnelling to neighbouring sites. We note
that this process can be made resonant for a site of lower energy, whilst far
detuned from a site of higher energy.

This is similar to the concept of sideband cooling in ion traps. There, the
goal is to cool the motional state of an atom in a single trap, by coupling to
states in which the electronic state is excited, but the motional state is
reduced by one quantum (this is called the red sideband). We can draw a figure
analogous to Fig.~\ref{Fig:compact}b, so that in comparison with sideband
cooling, we have replaced excited and ground electronic states with excited and
ground motional states in each well, and the coupling to lower motional states
in sideband cooling is replaced by coupling to neighbouring sites of lower
energy. In ion traps the goal is to cool an ion to the lowest motional state,
whereas in this proposal, sideband cooling leads to a spatial redistribution of
atoms towards the centre of the harmonic trap.

In analogy with sideband cooling, the cooling steps in this scheme (coupling to
sites of lower energy) will more strongly dominate heating steps (coupling to
sites of higher energy) when the energy offset between the sites is made
larger, and when the detuning $\delta \sim -\epsilon$, where $\epsilon$ is the
energy offset between sites. Note that in a harmonic trapping potential, where
the potential difference between neighbouring sites varies, the implementation
should involve a sweep in the detuning, making the cooling more efficient in
different parts of the trap as a function of time.

In Fig.~\ref{Fig:compact}c we plot example results for a single atom being
cooled in a harmonic trapping potential, obtained from monte-carlo wavefunction
simulations. The solid line in the figure shows the mean position of the atom
as a function of time for a single quantum trajectory, which illustrates the
cooling and heating transitions as the atom moves closer to and further from
the centre of the trap respectively. We also show the root-mean-square
displacement from the centre of the trap [i.e., $\sqrt{\langle(\sum_j j
\hat{n}_j)/M \rangle}$] when we begin with a uniform distribution for the
initial position. The finite final temperature is determined by competition
between heating and cooling steps, as is clearly illustrated by the transitions
in the example trajectory.

This method can be simply extended to fermions, where double occupation of the
ground motional state in any lattice site is prevented by Pauli blocking. For
this case it is possible to derive a QBME in analogy to that in sec.
\ref{Section:NParticles}, and to simulate these equations. Compaction of the
fermions into the centre of the trap is observed, producing a quantum register
with one fermion on each lattice site \cite{adunpublished}. An additional step
will, in general, be required to remove any remaining atoms from the upper
motional level at the end of the cooling process. In the present form, this
scheme is not well suited for bosons, as it would be difficult to prevent many
atoms from collecting on a single site, even in the presence of interactions.
However, the existence of this analogy is again a strong demonstration that
ideas from quantum optics can be used in the context of this new type of open
quantum system, an idea that has many possibilities for future exploration.

\section{Conclusions}\label{Section:Conclusion}

We have analysed a new cooling scheme based on ideas borrowed from sub-recoil
Raman cooling schemes, but where these ideas are now placed in the context of a
new form of open quantum system, where atoms in an optical lattice are coupled
to a BEC reservoir. In our case this setup provides laser assisted sympathetic
cooling, in which the final temperature of atoms in the lattice is not limited
by the temperature of the reservoir. This is motivated by the practical
requirement of achieving low temperatures in an optical lattice, which is
important for simulation of many strongly correlated quantum systems. Our
scheme indeed provides a possibility to achieve low temperatures, and from our
model we predict that temperatures a small fraction of the Bloch band width can
be achieved in this way.

Equally importantly, however, this work opens questions and possibilities on
the implementation of open quantum systems in this new context. Here we have
demonstrated how ideas from quantum optics can be applied in a many-body system
where we have strong control over many parameters, especially interactions
between atoms, and thus our system-reservoir coupling. The possibility exists
to extend these ideas to different types of dissipative Hubbard models, and
quantum reservoir engineering more generally. These ideas could be used both in
the context of state preparation and assist the engineering of new models with
systems of many atoms in optical lattices.

In the opposite direction, such open quantum systems also offer possibilities
for investigating effects that are of fundamental interest in quantum optics,
here in parameter regimes that can be very different from regular quantum
optics experiments. One example that is clear in the current context would be
possibility to study the analogue of superradiant or subradiant decay, using
the lattice and BEC parameters to engineer the wavelength of the excitations
created in the BEC, and therefore the strength of these effects.

\acknowledgments This work was supported by OLAQUI. Work in Innsbruck was
supported by the Austrian Science Foundation, the SCALA network IST- 15714, and
the Institute for Quantum Information; and work in Oxford by EPSRC project
EP/C51933/1.

\appendix
\section{Adiabatic elimination of the excited state}\label{Appendix:adiabatelim}

In this appendix we will show how the Raman coupling as given in eq.~(\ref{Ha}) can be derived
from eq.~(\ref{HRC0}) for large detuning $\Delta\gg \omega,\varepsilon_q^\alpha,\Omega$ of the two
lasers from the excited state and for a one dimensional optical lattice. Starting point are the
Heisenberg equations of motion for the field operators for the lattice atoms in ground and excited
state,
\begin{align}
    &\dot{\hat\psi}_a(x,t)=-\rmi\frac{\Omega_t(x)}{2}\hat\psi_e(x,t),\nonumber\\
    &\dot{\hat\psi}_e(x,t)=-\rmi\frac{\Omega_t^*(x)}{2}\hat\psi_a(x,t)+\rmi\Delta\hat\psi_e(x,t),\label{psidot}
\end{align} where we have introduced \begin{align}
    \Omega_t(x)=\Omega_1(x,t)+\rme^{-\rmi\delta t}\Omega_2(x,t).
\end{align}

Formal integration of eqs.~(\ref{psidot}) yields
\begin{align}
    \hat\psi_e(x,t)=-\frac{\rmi}{2}\int_0^t{\rm d}s\Omega_s^*(x)\rme^{\rmi\Delta(t-s)}\hat\psi_a(x,s),\label{psiet}
\end{align}
where we have assumed $\hat\psi_e(x,0)=0$ and $\Omega_{t<0}=0$. Expanding the field
operators in terms of Bloch wave functions gives \begin{align}
    \hat\psi_a(x,s)\approx \sum_{q,\alpha} \phi_q^\alpha(x)
    \rme^{\rmi(2\varepsilon_q^\alpha -\omega_\alpha)(t-s)}\hat A_q^\alpha(t),\label{psiat}
\end{align}
and inserting this into eq.~(\ref{psiet}) we find
\begin{align}
    \hat\psi_e(x,t)=&\sum_{q,\alpha}\phi_q^\alpha(x)
    \Big[\frac{1}{\Delta+2\varepsilon_q^\alpha-\omega_\alpha}
    \frac{\Omega_1^*(x,t)}{2}\nonumber\\ +&\frac{\rme^{\rmi\delta t}}
    {\Delta+2\varepsilon_q^\alpha-\omega_\alpha+\delta}\frac{\Omega_2^*(x,t)}{2}\Big]\hat A_q^\alpha(t),
\end{align}
after integrating the resulting equation, and we have assumed $\Omega_1(t)$, $\Omega_2(t)$ and $A_q^\alpha(t)$ to
be slowly varying on the timescale $1/\Delta$.

As the detuning $\Delta$ of the two Raman lasers from the excited state is the largest frequency,
i.e. $\Delta\gg \omega_\alpha,\delta,\varepsilon_q^\alpha$, we can neglect all terms rotating with
$\Delta$, and thus
\begin{align}
    \hat\psi_e(x,t)\approx \sum_{q,\alpha}\phi_q^\alpha(x)\left[\frac{\Omega_1^*(x,t)}{2\Delta}
    +\frac{\Omega_2^*(x,t)\rme^{\rmi\delta t}}{2\Delta}\right]\hat A_q^\alpha(t).
\end{align}
Inserting into the Heisenberg equations (\ref{psidot}) we can then read off the effective lattice
Hamiltonian in an interaction picture with respect to $H_0$ as
\begin{align}
    \hat H_{\rm a}&=\sum_{q,\alpha}\left(\frac{|\Omega_1|^2+|\Omega_2|^2}{4\Delta}(\hat A_q^\alpha)^\dagger \hat A_q^\alpha\right)
    \nonumber\\ &+\frac{\Omega}{2}\sum_{q,\alpha,\alpha'}R_{\alpha,\alpha'}(\dq)\Big(\rme^{\rmi(2
    \varepsilon_{q+\dq}^\alpha-2\varepsilon_q^\alpha+\omega_\alpha- \omega_{\alpha'} +
    \delta) t}\nonumber\\ &+
    \rme^{\rmi(2\varepsilon_{q+\dq}^\alpha-2\varepsilon_q^\alpha+\omega_\alpha- \omega_{\alpha'}-\delta) t}\Big)
    (\hat A_q^\alpha)^\dagger \hat A_{q+k,\alpha'}.
\end{align}
Here, we have assumed two running wave Raman lasers with relative momentum $\dq$ and neglected the
overlap of Wannier functions in different lattice sites. Transforming back to the Schr\"odinger
picture with respect to the lattice and only taking into account the lowest two Bloch bands we
obtain the Hamiltonian as given in eq.~(\ref{Ha}).

\section{Derivation of the master equation}\label{Appendix:ME}

In this section we will derive the master equation for the reduced system
density operator ${\hat\rho}$ for the decay of lattice atoms from the first
excited Bloch band back to the lowest band as given in eq.~(\ref{ME}) . As
described above, the BEC can be treated as a (three dimensional) $T=0$
reservoir, and the Born-Markov master equation in the interaction picture is
given by \cite{QN}
\begin{align}\label{ME0}
    \dot{\hat\rho}(t)=-\tr{\int_0^t dt'\Big[H_{\rm int}(t),[H_{\rm
    int}(t'),{\hat\rho}(t)\otimes{\hat\rho}_R]\Big]},
\end{align}
where ${\hat\rho}_R$ is the density operator for the BEC reservoir, and $\tr{}$
denotes the trace over the reservoir states. In a standard way we change to the
variable $\tau\equiv t-t'$ and extend the integration to $\tau \in [0,\infty)$,
assuming the correlation time in the BEC to be much shorter than typical system
time scales. Furthermore, we use $\int_0^\infty \rmd t\rme^{\rmi
(\varepsilon-\varepsilon_0) t} = \pi\delta(\epsilon-\epsilon_0)$ and
\begin{align}
  \EV{\hat b_\bk \hat b_{\bk'}^\dagger}=\delta_{\bk,\bk'} \quad \EV{\hat b_\bk^\dagger \hat
  b_{\bk'}}=0,
\end{align}
for an effective $T=0$ reservoir. In the rotating wave approximation (i.e. neglecting terms
rotating with frequencies $\omega$) we then find
\begin{align}
    \dot{\hat\rho}\approx \pi g_{ab}^2\frac{{\hat\rho}_b}{V}\sum_\bk&\left|\int\ddx w_{\bf 1}(\bx)w_0(\bx)\rme^{\rmi\bk\bx}\right|^2
    \delta(\omega - E_\bk) \nonumber\\
    &\left(2\hat c_\bk{\hat\rho} \hat c_\bk^\dagger - \hat c_\bk^\dagger \hat c_\bk{\hat\rho}-{\hat\rho} \hat c_\bk^\dagger
    \hat c_\bk\right),\label{MEint}
\end{align}
where we have furthermore utilized $J^0,|J^1|\ll\omega$ and neglected the terms
involving the band structure in the $\delta$ function. Note that we have also
neglected the energy shift terms resulting from principal value integrals in
this derivation. The energy shifts resulting from two atoms immersed in a 3D
reservoir have been computed to be very small \cite{kleinfleischhauer}.

In order to specialize to a one dimensional optical lattice in the three
dimensional reservoir we convert the sum in eq.~(\ref{MEint}) into an integral
in spherical coordinates, and integrate over $|\bk|$ (which is fixed by energy
conservation) and the azimuthal angle. For a one dimensional lattice we then
obtain
\begin{align}
    \dot{\hat\rho}=\sum_k \frac{\Gamma_k}{2} \left( 2\hat c_k{\hat\rho} \hat c_k^\dagger -\hat c_k^\dagger
    \hat c_k {\hat\rho} -{\hat\rho} \hat c_k^\dagger \hat c_k\right),
\end{align}
after converting the integral over the polar angle back to a sum over the
momentum in the $x$-direction, where $\Gamma_k$ is given by eq.~(\ref{GammaHO})
in harmonic oscillator approximation.

We can also rewrite the master equation in terms of position space operators,
$a^\alpha_i=(1/\sqrt{M}) \sum_q \exp(i q x_i) A_q^\alpha$. Defining $\tilde
c_i=(a_j^0)^\dag a_j^1$ We obtain
\begin{eqnarray}
    \dot\rho &=& \sum_k \frac{\Gamma_k}{2} \sum_{i\neq j} \left( 2\tilde c_i\rho \tilde c_j^\dagger -\tilde c_i^\dagger
    \tilde c_j \rho -\rho \tilde c_i^\dagger \tilde c_j\right) \rme^{-\rmi k (x_i-x_j)}\nonumber \\ & +& \sum_k \frac{\Gamma_k}{2} \sum_i \left( 2\tilde c_i\rho \tilde c_i^\dagger -\tilde c_i^\dagger
    \tilde c_i \rho -\rho \tilde c_i^\dagger \tilde c_i\right).
\end{eqnarray}
When $kd \gg 2\pi$, the terms in the sum over $i\neq j$ decay very rapidly with
$|i-j|$, as is shown in \cite{fermiloading} for the same master equation with a
fermionic reservoir. In this limit, collective effects of decay involving atoms
on different lattice sites can be neglected (i.e., there are no superradiant or
subradiant effects). See the discussion in section \ref{Section:ME} for more
details.

\section{Derivation of the quantum Boltzmann master equation}\label{Appendix:QBME}

In this section we will derive the quantum Boltzmann master equation (QBME),
eq.~(\ref{QBME}) which we use to analyze the dissipative coupling of lattice
atoms to the BEC reservoir in the case of many non-interacting bosons and
fermions, as presented in section \ref{Section:NParticles}. The QBME is an
equation for the diagonal elements $w_\bm=\bra{\bm} {\hat\rho} \ket{\bm}$ of
the system density operator ${\hat\rho}$ and can be derived from the master
equation (\ref{ME}). We project on the diagonal elements of the density
operator and find
\begin{align}
    &\dot w_{\bm}=\sum_k\frac{\Gamma_k}{2}\sum_{\bn,q,q'} \nonumber\\
    &\Big(2\bram (A_{q-k}^0)^\dagger A_{q}^1\ketn\bran
    (A_{q'}^1)^\dagger A_{q'-k}^0\ketm w_{\bn} \label{Appendix:QBME0_1}\\
    &-\bram (A_{q}^1)^\dagger A_{q-k}^0 (A_{q'-k}^0)^\dagger A_{q'}^1\ketn\bran\bm\rangle w_{\bn}\label{Appendix:QBME0_2}\\
    &-w_{\bn} \bram\bn\rangle\bran (A_{q}^1)^\dagger A_{q-k}^0
    (A_{q'-k}^0)^\dagger A_{q'}^1\ketm\Big).\label{Appendix:QBME0_3}
\end{align}
The expectation values in the first part (\ref{Appendix:QBME0_1}) of this equation can be
calculated as
\begin{align}
    &\bran (A_{q'}^1)^\dagger A_{q'-k}^0\ketm = \sqrt{m_{q'-k}^0} \sqrt{1\pm
    m_{q'}^1} \delta_{\bm,\bn+\be_{q'-k,q'}},\nonumber \\
    &\bram (A_{q-k}^0)^\dagger A_{q}^1\ketn = \sqrt{m_{q-k}^0} \sqrt{1\pm
    m_{q}^1} \delta_{\bm,\bn+\be_{q-k,q}},
\end{align}
where the upper (lower) signs are for bosons (fermions), the product of the two $\delta$ functions
appearing in these expressions can be evaluated as $\delta_{\bm,\bn+\be_{q'-k,q'}}
\delta_{\bm,\bn+\be_{q-k,q}}= \delta_{\bm,\bn+\be_{q-k,q}} \delta_{q,q'}$. For the second and
third line (\ref{Appendix:QBME0_2}) and (\ref{Appendix:QBME0_3}) we find the identical expressions
\begin{align}
    &\bram (A_{q}^1)^\dagger A_{q-k}^0 (A_{q'-k}^0)^\dagger
    A_{q'}^1\ketn\bran\bm\rangle= \nonumber \\
    &m_q^1 (1\pm m_{q-k}^0)\delta_{q,q'}
    \delta_{\bm,\bn}.
\end{align}
Thus, we can perform the sums over $q'$ and $\bn$ and end up with the QBME
\begin{align}
    \dot{w}_{\bm}=&\sum_{k,q}\Gamma_k\left[ \rmm_{q-k}^0(1\pm\rmm_{q}^1)w_{\bm'}
    - \rmm_{q}^1 (1\pm\rmm_{q-k}^0)w_{\bm}\right],
\end{align}
with $\bm'=\bm-\be_{q-k,q}$.

\end{document}